\documentclass[oldversion]{aa}  
\usepackage{graphicx,float}
\usepackage{latexsym,amsmath,amssymb}

\usepackage{txfonts}

\usepackage{epstopdf}

\begin{document}
   \title{Reflection in Seyfert galaxies and the unified model of AGN}

   \subtitle{}

   \author{C. Ricci
          \inst{1,2},
          R. Walter
          \inst{1,2},
          T. J.-L. Courvoisier
          \inst{1,2}, 
           \and
           S. Paltani \inst{1,2} 
          }

   \institute{ \textsl{ISDC} Data Centre for Astrophysics, University of Geneva, ch. d'Ecogia 16, 1290 Versoix, Switzerland 
    \and Geneva Observatory, University of Geneva, ch. des Maillettes 51, 1290 Versoix, Switzerland \\
             }
    \offprints{e-mail: Claudio.Ricci@unige.ch} 
   \authorrunning{ C. Ricci et al.}
   \titlerunning{Reflection in Seyfert galaxies and the unified model of AGN}
    \date{Received; accepted}

 \abstract{
We present a deep study of the average hard X-ray spectra of Seyfert galaxies. We aim to test the unified model of active galactic nuclei, and constrain differences and similarities between different classes of objects. 
 We analyzed all public \textit{INTEGRAL} IBIS/ISGRI data available on all the 165 Seyfert galaxies detected at $z<0.2$. Our final sample consists of 44 Seyfert\,1s, 29 Seyfert\,1.5s, 78 Seyfert\,2s, and 14 narrow-line Seyfert\,1s. For each subsample, we stacked all the images, and derived their average hard X-ray spectra in the 17--250 keV energy range. We performed a detailed spectral analysis using both a model-independent and a model-dependent approach.
  All classes of Seyfert galaxies show on average the same nuclear continuum, as foreseen by the zeroth order unified model, with a cutoff energy of $E_C\gtrsim200$\,keV, and a photon index of $\Gamma\simeq 1.8$. The average optical depth of the Comptonizing medium is consistent for the different classes ($\tau \simeq 0.8$). Compton-thin Seyfert\,2s show a reflection component stronger than Seyfert\,1s and Seyfert\,1.5s. Most of this reflection is due to mildly obscured ($10^{23}\rm \,cm^{-2} \leq N_{\rm \,H} < 10^{24}\rm \,cm^{-2}$) Seyfert\,2s, which have a significantly stronger reflection component ($R=2.2^{+4.5}_{-1.1}$) than Seyfert\,1s ($R\leq 0.4$), Seyfert\,1.5s ($R\leq 0.4$) and lightly obscured ($N_{\rm \,H} < 10^{23}\rm \,cm^{-2}$) Seyfert\,2s ($R\leq 0.5$). This cannot be explained easily by the unified model. The absorber/reflector in mildly obscured Seyfert\,2s might cover a large fraction of the X-ray source, and contain clumps of Compton-thick material. The large reflection found in the spectrum of mildly obscured Seyfert\,2s reduces the amount of Compton-thick objects needed to explain the peak of the cosmic X-ray background. Our results are consistent with the fraction of Compton-thick sources being $\sim 10\%$. The spectra of Seyfert\,2s with and without polarized broad lines do not show significant differences, the only difference between the two samples being the higher hard X-ray and bolometric luminosity of Seyfert\,2s with polarized broad lines. The average hard X-ray spectrum of narrow-line Seyfert\,1s is steeper than those of Seyfert\,1s and Seyfert\,1.5s, probably due to a lower energy of the cutoff.
}

   \maketitle
   
\section{Introduction}
Active Galactic Nuclei (AGN) emit over the entire electromagnetic spectrum, and are commonly assumed to be powered by accretion onto a super massive black hole (Rees 1984).
Seyfert galaxies host AGN, and are classified according to their emission lines as Seyfert\,1s (Sy1s, showing broad and narrow emission lines), and Seyfert\,2s (Sy2s, showing only narrow lines). Many Seyfert galaxies exhibit optical spectra with properties in-between those of Sy1 and Sy2 galaxies (e.g., Osterbrock \& Koski 1976). These objects have been classified as type 1.2, 1.5, 1.8, or 1.9 Seyfert galaxies, depending on the details of their optical spectra.
Studies of the optical spectra of Seyfert\,2 galaxies using polarized light (Miller \& Antonucci 1983, Antonucci \& Miller 1985) showed that broad emission lines can also be detected in some of these objects, and led to the development of the so-called unified model (UM, Antonucci, 1993).
According to this model, the same engine is at work in all kind of Seyfert galaxies, and the differences between Seyfert\,1s and Seyfert\,2s can be ascribed solely to orientation effects and to anisotropic obscuration. In this model, the line of sight to the nucleus is (Seyfert\,2s) or is not (Seyfert\,1s) obstructed by optically thick material, possibly distributed in a toroidal geometry. X-ray observations have confirmed this idea, showing that most AGN unabsorbed in X-rays are of the optical Seyfert~1 type, and that most AGN that are absorbed belong to the Seyfert~2 group (e.g., Awaki et al. 1991). 

However, in the past few years observational evidence of significant differences between Seyfert\,1s and Seyfert\,2s has been discovered. Seyfert\,1s with significant absorption have been found (e.g., Fiore et al. 2001, Cappi et al. 2006), along with Seyfert\,2s without X-ray absorption (e.g., Pappa et al. 2001, Panessa \& Bassani 2002).
Furthermore, spectropolarimetric surveys indicate that only $\sim 30-50\%$ of Seyfert\,2s have polarized broad lines (PBLs), which might imply that not all of them harbor hidden BLRs (e.g., Tran 2001, 2003, Gu \& Huang 2002). In the soft X-rays the photon index distribution is found to be skewed towards low values of $\Gamma$ for obscured objects (Brightman \& Nandra 2010), while in the hard X-rays, Seyfert\,2s have been found to have harder spectra than Seyfert\,1s (e.g., Zdziarski et al. 1995, Malizia et al. 2003, Deluit \& Courvoisier 2003, Ajello et al. 2008a, Burlon et al. 2011).

Two subclasses of objects cannot be easily explained by the unified model: narrow-line Seyfert\,1s (NLS1s) and low ionization nuclear emission line regions (LINERs). 
It has been suggested that NLS1s are AGN in their early phase (Grupe et al. 1999), characterized by relatively low black hole masses (e.g., Grupe \& Mathur 2004) and very high accretion rates in terms of Eddington units (e.g., Grupe et al. 2010). LINERs (Heckman 1980) are low luminosity AGN, and display multiwavelength peculiar characteristics. They could be the link between AGN and normal galaxies (Zhang et al. 2009, Rovilos et al. 2009), or their SMBH might accrete differently from normal Seyfert galaxies (Ho 2008).

The X-ray emission of Seyfert galaxies is thought to be produced by the Comptonization of ultraviolet (UV) photons generated in the innermost edge of the accretion disk by a population of hot electrons located in a coronal region sandwiching the disk (Haardt \& Maraschi 1991, 1993). The X-ray spectrum of AGN can be normally well described by a power law model with a photon index between $\Gamma \simeq 1.8$ (e.g., Dadina et al. 2008) and $\Gamma \simeq 2$ (e.g., Beckmann et al. 2009). Other common characteristics are an exponential cutoff at an energy depending on the temperature of the plasma $T_e$ and its Thomson opacity, and photoelectric absorption at low energies, caused by material along the line of sight. Two prominent features generated by the reflection of the continuum are also observed in the X-ray spectra of AGN: a neutral iron K$\alpha$ line (at 6.4 keV in the local reference frame) and a reflection hump peaking at E~$\simeq 30$ keV (Magdziarz \& Zdziarski 1995). At energies $\gtrsim\,20$\,keV, the photoelectric cross-section sharply declines, and the presence of matter along the line of sight does not play a significant role, unless the source is Compton-thick (CT; $N_{\rm \,H} > 1.5 \times 10^{24} \rm \,cm^{-2}$). 
Therefore hard X-rays are ideal for probing the intrinsic emission of AGN and testing the UM, according to which different types of Seyfert galaxies should show on average the same characteristics.

The {\it INTErnational Gamma-Ray Astrophysics Laboratory} ({\it INTEGRAL}; Winkler et al. 2003) is a hard X-ray/soft $\gamma$-ray
mission, designed for imaging and spectroscopy with high angular and spectral resolution in the energy range from $\sim 3 \rm$ \,keV to
10 MeV.  Since its launch on October 17, 2002, the soft $\gamma$-ray imager IBIS/ISGRI (15--1000 keV; Lebrun et al. 2003, Ubertini et al. 2003) on board {\it INTEGRAL} has detected more than 200 AGN, among which about 90 were detected for the first time in the hard X-rays. IBIS/ISGRI uses the coded-aperture technique (Caroli et al. 1987) and has a large field of view of 29\,$\degr$ square with a spatial resolution of 12\,arcmin.

We present a very deep study of the average hard X-ray spectra of Seyfert galaxies.
Our main aim is to test the unified model by studying the average hard X-ray emission of different types of radio-quiet AGN detected by {\it INTEGRAL} IBIS/ISGRI, and to constrain the average spectral characteristics of each class. We also investigate whether Seyfert\,2s showing (hereafter PBL Sy2s) and not showing (hereafter NPBL Sy2s) PBLs are intrinsically similar objects or not.
In Sect.\,\ref{Sample}, we present the sample of AGN; in Sect.\,\ref{sec:integral}, we describe how the data analysis was performed and how the hard X-ray spectra were obtained; in Sect.\,\ref{sect:modIndep}, we present the results obtained from a model-independent analysis of the spectra, and in Sect.\,\ref{spectral_analysis}, we discuss the model-dependent spectral analysis. We study the origin of a strong reflection component found in the spectrum of Seyfert\,2s in Sect.\,\ref{sect:bump}; and in Sect.\,\ref{sect:discussion}, we discuss our results and their implications for the UM and the cosmic X-ray background (CXB). In Sect.\,\ref{sect:concl}, we present our conclusions and summarize our findings.

\section{The sample}
\label{Sample}
The sample consists of all the 205 Seyfert galaxies detected (significance $> 5 \sigma $) in the hard X-rays by {\it INTEGRAL} IBIS/ISGRI during its first eight years of operations. 
We considered as their optical classifications the ones  reported in the latest version of the Veron Catalog of Quasars and AGN (Veron-Cetty \& Veron, 2010).

The sample consists of 38 Seyfert\,1s, 17 Seyfert\,1.2s, 34 Seyfert\,1.5s, 3 Seyfert\,1.8s, 15 Seyfert\,1.9s, 75 Seyfert\,2s, 8 LINERs, and 15 NLS1s. We excluded from these subsamples the LINERs, because the physical characteristics of their accretion flow might differ from those of the others. We grouped Seyfert\,1.2s with Seyfert\,1s, and both Seyfert\,1.8s and Seyfert\,1.9s with Seyfert\,2s because of their similarity.

\subsection{The role of absorption}

\begin{figure}[h]
\centering
\includegraphics[width=9cm]{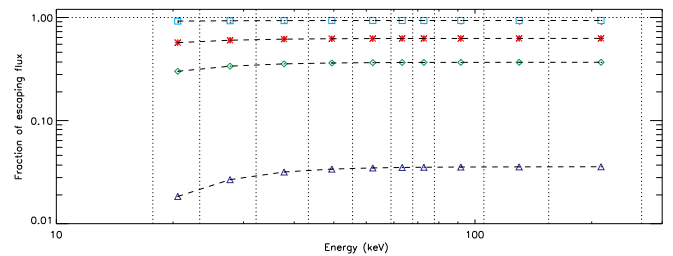}
\caption{Fraction of escaping flux in the 10 energy bins used. Squares represent $N_{\rm \,H}=10^{23} \rm \,cm^{-2}$, stars $N_{\rm \,H}=7\times 10^{23} \rm \,cm^{-2}$, diamonds $N_{\rm \,H}=1.5\times 10^{24} \rm \,cm^{-2}$, and triangles $N_{\rm \,H}=5\times10^{24} \rm \,cm^{-2}$.}
\label{fig:escapingflux}
\end{figure}

To model the effect of absorption on the X-ray spectra of AGN, one has to take into account both photoelectric absorption and Compton scattering.
The photoelectric cross-section $\sigma_{\rm\,ph}$ has a strong dependence on the energy, while Compton scattering in the hard X-rays depends on the Thomson cross-section $\sigma_{\rm \,T}$, and is constant with energy up to the Klein-Nishina decline.
The cumulative effect of the two cross-sections is given by
\begin{equation}\label{eq:abs}
M(E)=e^{-\sigma_{\rm\,ph}(E)N_{\rm \,H}}\times e^{-\sigma_{\rm \,T}N_{\rm \,H}}.
\end{equation}
For column densities of the order of $\sigma_{\rm \,T}^{-1}=1.5\times 10^{24} \rm \,cm^{-2}$, only about 30\% of the flux is able to escape at $\sim\,20\,$keV (see Fig.\,\ref{fig:escapingflux}). About 60\% of the original emission is unabsorbed for $N_{\rm \,H}=7\times10^{23}\rm \,cm^{-2}$, and we used this value as a threshold between Compton-thin and Compton-thick sources. Thus, we included in our final sample of Seyfert\,2s only the objects with a value of column density of $N_{\rm \,H}\,\leq\,7\times10^{23}\rm \,cm^{-2}$, and placed the 10 Seyfert\,2s with $N_{\rm \,H}\,>\,7\times10^{23}\rm \,cm^{-2}$ in the CT Seyfert\,2s sample. We note, however, that none of the sources in our sample has a $N_{\rm\,H}$ in the (7--10)$\times 10^{23} \rm\,cm^{-2}$ range, so in the following we refer to $10^{24} \rm\,cm^{-2}$ as the threshold between Compton-thin and Compton-thick objects.
In the following, we use solar abundances and the photoelectric cross section of Morrison and McCammon (1983).

\begin{table}
\caption[]{Sources detected by {\it INTEGRAL} IBIS/ISGRI but excluded from the analysis, and reason for their exclusion. More details can be found in Sect.\,\ref{sec:excludesources}. }
\label{tab:sourcesexcluded}
\begin{center}
\begin{tabular}{llc}
\hline
\hline
\noalign{\smallskip}
Source & Type & Reason for the exclusion \\ 
\noalign{\smallskip}
\hline
\noalign{\smallskip}
4U 1344$-$60         &  Sy 1.5  &   {\it C}  \\                    		
Circinus Galaxy       &  CT Sy 2  &  {\it B} \\                       
ESO 138$-$1        &  Sy 2  &  {\it C}   \\                     		
ESO 383$-$18	       &  Sy 2  &  {\it C}   \\ 
GRS 1734$-$292        &  Sy 1  &  {\it B}    \\                        
IC 4329A 		    &  Sy 1  &  {\it B}   \\   			
IGR J00465$-$4005       &  Sy 2  &  {\it R}    \\ 		
IGR J05270$-$6631       &  Sy 1  &  {\it R}    \\ 		
IGR J06117$-$6625       &  Sy 1.5  &  {\it R}    \\   		 
IGR J09523$-$6231       &  Sy 2  &  {\it R}    \\ 	
IGR J10147$-$6354       &  Sy 1  &  {\it R}    \\		
IGR J12131+0700     & Sy 1.5   &  {\it C}   \\        
IGR J13038+5348       &  Sy 1  &  {\it R}    \\  
IGR J14515$-$5542        & Sy 2   &   {\it C}  \\                   
IGR J16024$-$6107       & Sy 2   &  {\it C}   \\                    
IGR J16056$-$6110       & Sy 1.5   &  {\it C}   \\                 
IGR J16426+6536 	       &  NLS1  &  {\it U}   \\
IGR J17476$-$2253        &  Sy 1  &  {\it C}   \\ 
IGR J17488$-$3253       &  Sy 1  &  {\it C}   \\ 
IGR J18249$-$3243       &  Sy 1  &  {\it R}    \\	
IGR J21272+4241        & Sy 2 & {\it R} \\	
NGC 1365       &  Sy 2  &  {\it U}    \\
NGC 4151     &  Sy 1.5  &  {\it B}    \\                        	
NGC 4388        &  Sy 2  &  {\it B}    \\                        	
NGC 4939       &  Sy 2  &  {\it U}    \\
NGC 6221       &  Sy 2  &  {\it C}   \\    
PKS 0637$-$752       &  Sy 1.5  &  {\it R}    \\ 			
QSO B1821+643       &  Sy 1  &  {\it R}    \\ 		
SWIFT J0216.3+5128       &  Sy 2  &  {\it R}    \\ 
Was 49a + Was 49b 	       &  Sy 1 + Sy 2  &  {\it U}   \\ 
\noalign{\smallskip}
\hline
\noalign{\smallskip}
\multicolumn{3}{l}{{\bf Notes.} {\footnotesize {\it C}: close to other detected sources ($\leq0.4 ^{\circ}$); {\it U}: classification}} \\
\multicolumn{3}{l}{{\footnotesize unclear or with particular characteristics; {\it R}: $z > 0.2$; {\it B}: bright}} \\
\multicolumn{3}{l}{{\footnotesize (detection significance $> 50\sigma$ in the 17--80 keV band).}} \\
\end{tabular}
\end{center}
\end{table}
\begin{figure}[t!]
\centering
\includegraphics[width=9cm]{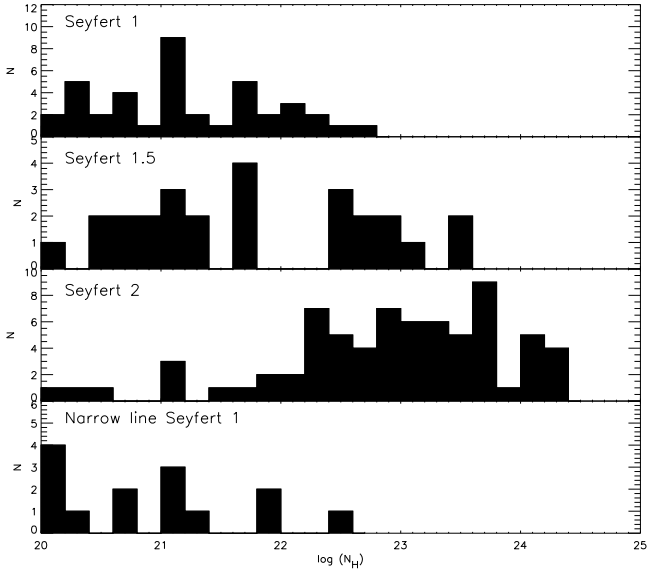}
\caption{Hydrogen column density distributions of Seyfert\,1s, Seyfert\,1.5s, Seyfert\,2s and narrow-line Seyfert\,1s.}
\label{fig:NHhist}
\end{figure}%
\subsection{Excluded sources}\label{sec:excludesources}
To avoid contamination and problems related to the way we obtain the average spectra (see Sect.\,\ref{sec:integral}), we excluded from the sample 10 sources which are too close ($\le\,0.4^{\circ}$) to each other or to other bright sources. We excluded 11 sources with redshift  $z>\,0.2$ to avoid spurious effects due to redshifted spectral features (i.e. the high-energy cutoff and the reflection hump). We removed the 5 sources detected with a significance $>$\,$50\sigma$, to prevent them from dominating the final spectra.

Four sources with either an uncertain classification or peculiar absorbers were also excluded from the analysis. Was\,49 is an interacting system of galaxies containing two Seyfert nuclei (Moran et al. 1992). The AGN IGR\,J16426+6536 shows characteristics typical of both NLS1 (Masetti et al. 2009) and Seyfert\,1.5 (Butler et al. 2009) galaxies.
The Seyfert\,1.8 NGC\,1365 has been found to undergo transitions from a ``transmission dominated" to a ``reprocessing dominated" state and back in a few weeks time, which have been interpreted as due to eclipses of CT material along the line of sight (Risaliti et al. 2005). A similar state transition has been observed in NGC\,4939 (Guainazzi et al. 2005), a Seyfert\,2 that was previously classified as Compton-thick by {\it BeppoSAX} observations (Maiolino et al. 1998).
The excluded sources are reported in Table\,\ref{tab:sourcesexcluded}.

\subsection{The final sample}\label{sec:finsample}
The final sample used for the analysis consists of 165 AGN, of which 44 are Seyfert\,1s, 29 Seyfert\,1.5s, 68 Compton-thin Seyfert\,2s, 10 CT Seyfert\,2s, and 14 NLS1s.
Spectropolarimetric data are available for only 19 of the 68 Compton-thin Seyfert\,2s of our sample. Among these, 11 show and 8 do not show PBLs. The sources and their characteristics are reported in Appendix\,\ref{list_sources}.

Figure\,\ref{fig:NHhist} shows the distribution of the column density for the different classes of AGN. The average values of the column densities of the different classes are $N_{\rm \,H}^{\rm \,Sy1}=5.4\times10^{21}\rm \,cm^{-2}$, $N_{\rm \,H}^{\rm \,Sy1.5}=3.7\times10^{22}\rm \,cm^{-2}$, $N_{\rm \,H}^{\rm \,Sy2}=1.5\times10^{23}\rm \,cm^{-2}$, $N_{\rm \,H}^{\rm \,NLS1}=1.9\times10^{21}\rm \,cm^{-2}$, and $N_{\rm \,H}^{\rm \,CT}=1.5\times10^{24}\rm \,cm^{-2}$, for Seyfert\,1s, Seyfert\,1.5s, Seyfert\,2s, NLS1s, and CT Seyfert\,2s, respectively. These values agree with the paradigm that Seyfert\,2s are more obscured than Seyfert\,1s and Seyfert\,1.5s. Considering the average column densities of the different samples, the percentage of flux absorbed in the first band (17--22 keV) by photoelectric processes is about 0.01\%, 0.5\%, and 2\% for Seyfert\,1s, Seyfert\,1.5s, and Seyfert\,2s, respectively. Compton processes have a stronger influence at these energies, and the fraction of scattered flux is of 0.3\%,  2.4\% and 9.5\%, for Seyfert\,1s, Seyfert\,1.5s and Seyfert\,2s, respectively. In the case of CT Seyfert\,2s, the influence of the obscuring material is stronger and $\simeq 18 \%$ of the flux is photoelectrically absorbed, whereas when we also consider Compton scattering, only $\simeq 30\%$ of the flux escapes. A Kolmogorov-Smirnov (KS) gives a probability of $\simeq\,67\%$ that the column density distributions of Seyfert\,1s and Seyfert\,1.5s are drawn from the same parent population. The probability is much lower when comparing NLS1s to Seyfert\,1s and Seyfert\,1.5s ($\simeq\,10\%$).

\begin{figure}[t!]
\centering
\includegraphics[width=9cm]{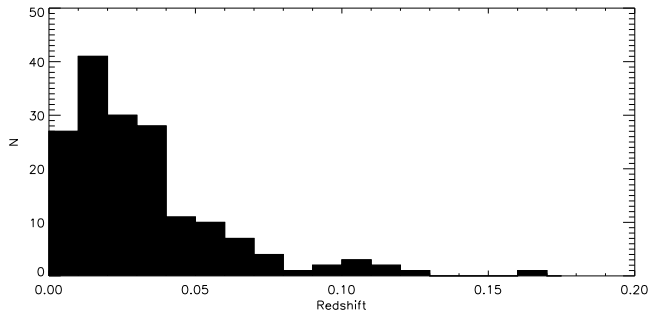}
\caption{Redshift distribution of our final sample of \textit{INTEGRAL} IBIS/ISGRI detected Seyfert galaxies.}
\label{fig:Redshift}
\end{figure}

In Fig.\,\ref{fig:Redshift}, we show the redshift distribution of our final sample. The average value of the redshift is z=0.03, and the average detection significance is 12.7$\sigma$.

\begin{figure}[t!]
\centering
\includegraphics[width=9cm]{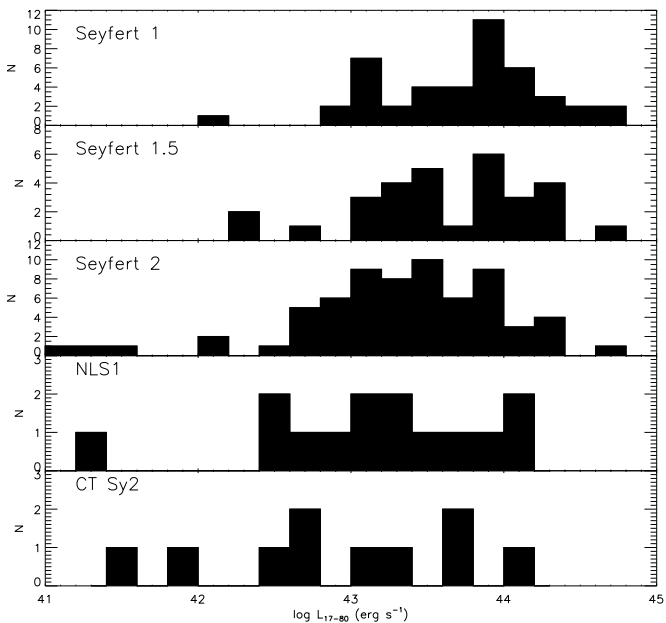}
\caption{Luminosity distribution of Seyfert\,1s, Seyfert\,1.5s, Compton-thin Seyfert\,2s, NLS1s, and CT Sy2s in the 17--80 keV energy band.}
\label{fig:Lumin}
\end{figure}

In Fig. \ref{fig:Lumin}, we show the luminosity distributions in the 17--80 keV band of our samples of Seyfert galaxies. The luminosities were calculated assuming a standard $\Lambda$CDM cosmological model, Crab-like spectra (e.g., Jourdain \& Roques 2009), and $H_0=70 \rm \,km \,s^{-1} \,Mpc^{-1}$. The average values of the luminosity are $L_{\rm \,Sy1}=5.0\times 10^{43}\rm \,erg \,s^{-1}$, $L_{\rm \,Sy1.5}=3.7\times 10^{43}\rm \,erg \,s^{-1}$, $L_{\rm \,Sy2}=2.1\times 10^{43}\rm \,erg \,s^{-1}$, $L_{\rm \,CT Sy2}=8.3\times 10^{42}\rm \,erg \,s^{-1}$, and $L_{\rm \,NLS1}=1.7\times 10^{43}\rm \,erg \,s^{-1}$, for Sy1s, Sy1.5s, Sy2s, CT Sy2s, and NLS1s, respectively. The luminosity of CT Sy2 was corrected for Compton scattering. A KS test gives a probability of $\simeq 53\%$ that the luminosity distributions of Seyfert\,1s and Seyfert\,1.5s are statistically compatible. The probability is much lower when comparing Seyfert\,1s to Seyfert\,2s ($\simeq 2\%$), and Seyfert\,1s to NLS1s ($\simeq10\%$).

\section{\textit {INTEGRAL} IBIS/ISGRI data analysis}
\label{sec:integral}
We used all the public data obtained by {\it INTEGRAL} IBIS/ISGRI as of May 2010, for a total of more than 50 thousands pointings or ``science windows'' (SCWs). The typical exposure time of these pointings is of $(1-3)\times 10^3$ s. Several thousand pointings, including at least one source of the sample in its field of view and with an effective exposure longer than 120 s were selected, spanning times between December~30, 2002 (revolution 26) and April~7, 2009 (revolution 791).

The {\it ISGRI} data were reduced using the {\it INTEGRAL}  Offline Scientific Analysis software\footnote{\texttt{http://www.isdc.unige.ch/}} version 9.0, publicly released by the {\it ISDC} Data Centre for Astrophysics (Courvoisier et al. 2003). 

Individual sky images for each pointing were produced in a broad energy band (17--250~keV), and divided into ten bins as follows: 17--22, 22--30, 30--40, 40--51, 51--63, 63--71, 71--80, 80--105, 105--150 , and 150--250 keV.

To extract the average spectra of the different AGN samples, we followed the procedure adopted by Walter \& Cabral (2009). We created 500$\times$500-pixels mosaic images modifying the coordinate system of each individual image, setting the coordinates of each source of the sample to an arbitrary fixed position ($\alpha$=0, $\delta$=0). The geometry of the image was also modified to have a consistent PSF independently of the position of the source in the field of view (FOV). These mosaic images provide a stack of all the selected IBIS/ISGRI data for each considered sample.

To minimize the systematics in the mosaic images, we excluded 1455 individual sky images (many obtained before revolution 38, when the IBIS bottom anti-coincidence was reconfigured) that had a background fluctuation rms larger than $1.1\sigma$ in the significance image, and 779 images with a minimum significance smaller than $-5.5\sigma$. In addition to these, we excluded 627 images taken when the ISGRI was in staring mode. A total of 56,611 images were finally included in the processing (some of them many times, when including several sources). All known ISGRI sources with significance above $5\sigma$ were used for image cleaning. When known accurately, the position of these sources was fixed to catalogue values taken from the \textit{INTEGRAL} general reference catalog (Ebisawa et al. 2003)\footnote{\texttt{http://www.isdc.unige.ch/integral/science/catalogue}}. 

The mosaic images were built with a tangential projection using a factor of two oversampling when compared to the individual input sky images, this results in a pixel size of 2.4 arcmin at the center of the mosaic. The photometric integrity and accurate astrometry were obtained by calculating the intersection between input and output pixels, and weighting the count rates according to the overlapping area. 

The average signal extracted in the large 17--80 keV band from each ISGRI mosaic, and both the exposures and the number of sources used are reported in Table\,\ref{tab:flux}. 
The detection significances in the 17--80 keV band range between 102.8$\sigma$ and 32.9$\sigma$, for the largest and smallest sample (i.e. Seyfert\,2s and CT Sy2s), respectively. The effective exposure obtained at the center of the mosaics are between 4.7\,Ms (for CT Sy2s) and 44.0\,Ms (for Seyfert\,2s).

The ten-bin spectra of the final samples were extracted from the images in the 10 narrow bands using {\tt mosaic\_spec}. In all spectra, the highest energy bin (i.e. 150--250 keV) has a low significance ($\sim 2.5-3\sigma$), but we use it for the sake of completeness. We used the latest detector response matrix (RMF), and calculated the ancillary response matrices (ARFs) of each sample with a weighted average of the nine available ARFs, based on the number of SCWs within the validity time of a particular ARF.

\begin{table}[t]
\caption{Number of sources (1), effective exposures (2), detection significances (3), and count rates (4) of the different samples. All the values refer to the 17--80 keV band.
}
\label{tab:flux}
\resizebox{\columnwidth}{!}{%
\begin{tabular}[c]{lcccc}
\hline \hline \noalign{\smallskip}
 & (1) & (2) & (3) & (4) \\
Sample&Srcs&Exp.& Det. Significance&Ct rate\\
               &&[{\tiny Ms}]&  [{\tiny $\sigma$}] &[{\tiny ct/s}]\\
\noalign{\smallskip\hrule\smallskip}
Seyfert\,1      & 44 &  37.6   &  91.1  & $0.396 \pm 0.004 $ \\
Seyfert\,1.5        & 29  &  13.2  & 59.1  & $ 0.429 \pm 0.007 $ \\
Seyfert\,2      &  68  &  44.0  &  102.8  & $ 0.390 \pm 0.004 $ \\
CT Sy\,2      &  10 &   4.7 &  32.9  & $  0.405  \pm 0.012  $ \\
NLS1         &14& 10.1   & 44.3 &$0.377 \pm 0.008$ \\      
\noalign{\smallskip\hrule\smallskip}
MOB Sy2         &  27 &  13.8 & 59.6  &$  0.430 \pm 0.007 $ \\      
LOB Sy2        & 34 & 23.4 & 78.1  &$   0.424 \pm 0.006 $ \\
\noalign{\smallskip\hrule\smallskip}
PBL Sy2         & 11 & 4.2  & 99.0  &$ 1.301 \pm 0.013 $ \\      
NPBL Sy2        & 8 & 2.9  & 31.2  &$ 0.479 \pm 0.015 $ \\     
\noalign{\smallskip\hrule\smallskip}
\end{tabular}
}
\end{table}

\section{Model-independent spectral analysis}\label{sect:modIndep}
A model-dependent spectral analysis has the drawback of possible parameter degeneracy, which might not allow us to constrain several parameters at the same time. An alternative method to characterize differences and similarities between different classes of Seyfert galaxies, independently of their average flux, is by means of a model-independent approach. This was achieved by normalizing the flux of the two different spectra in the first bin (17--22 keV), and then calculating their ratio.

\subsection{Seyfert galaxies}

We have no information about the value of the hydrogen column density of seven Seyfert\,2s. To discard the possibility that these objects are CT, which might influence significantly the result, we calculated the ratio between the spectra of the Seyfert\,2 sample including and excluding these seven objects. This ratio is fully consistent with 1.

\begin{figure}[h]
\centering
\includegraphics[width=9cm]{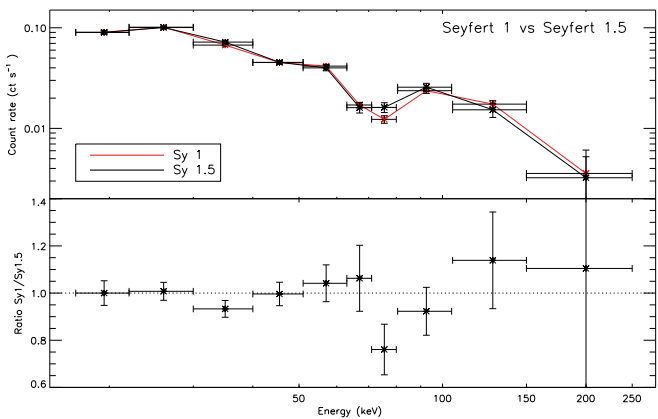}
\caption{Normalized spectra ({\it upper panel}) and ratio ({\it lower panel}) of Seyfert\,1s to Seyfert\,1.5s.}
\label{fig:sy1vs1.5}
\end{figure}

\begin{figure}[h]
\centering
\includegraphics[width=9cm]{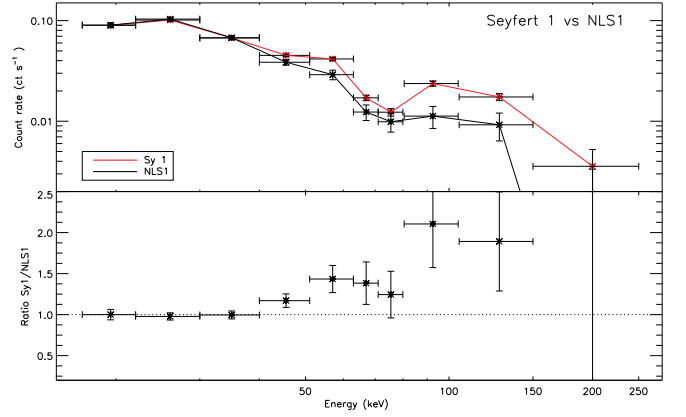}
\caption{Normalized spectra ({\it upper panel}) and ratio ({\it lower panel}) of Seyfert\,1s to NLS1s.}
\label{fig:sy1vsNLS1}
\end{figure}

\begin{figure}[h]
\centering
\includegraphics[width=9cm]{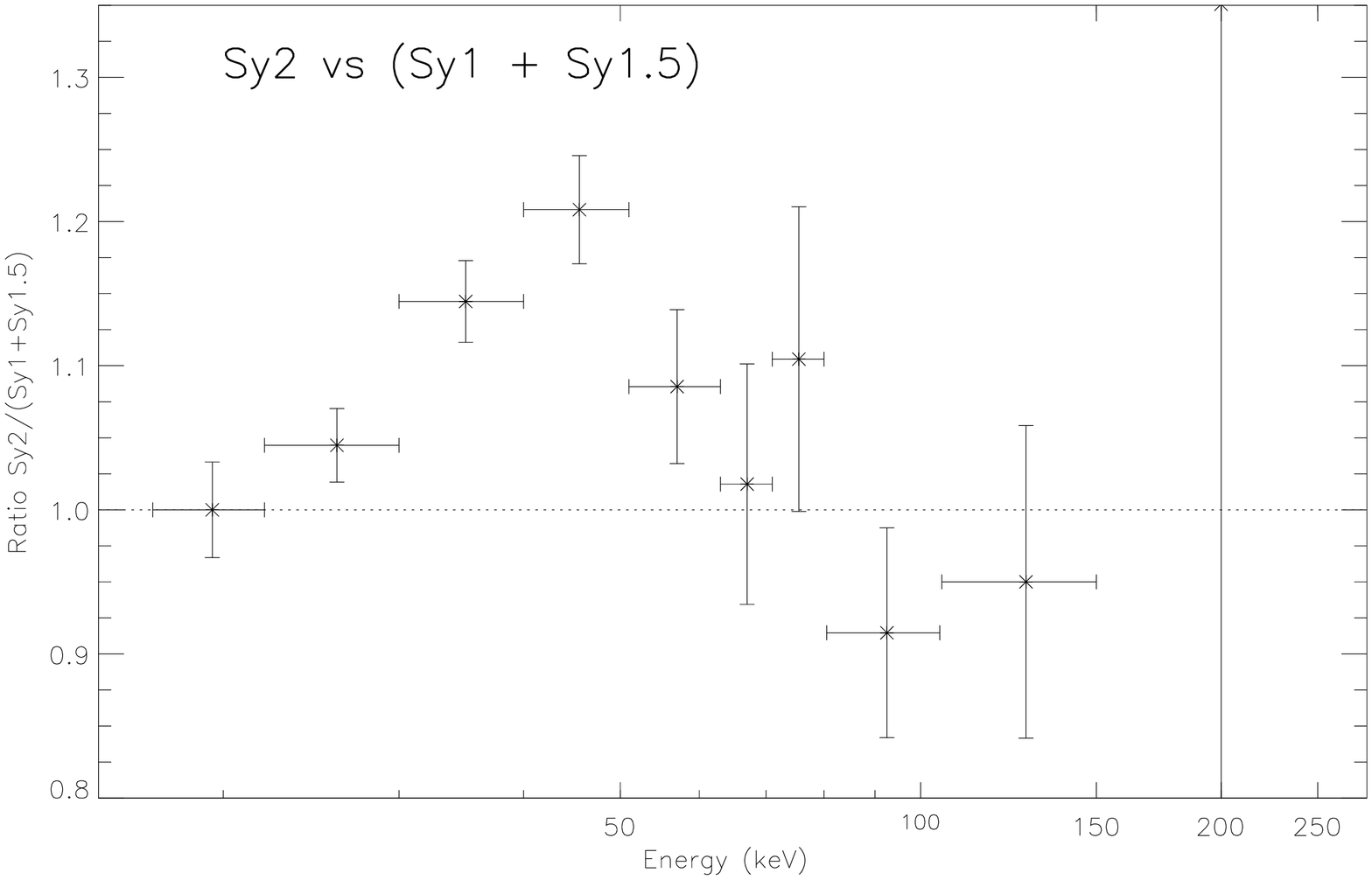}
\caption{Ratio between the normalized spectra of Compton-thin Seyfert~2s and that of both Seyfert~1s and Seyfert~1.5s.}
\label{fig:merged}
\end{figure}

\begin{figure}[h]
\centering
\includegraphics[width=9cm]{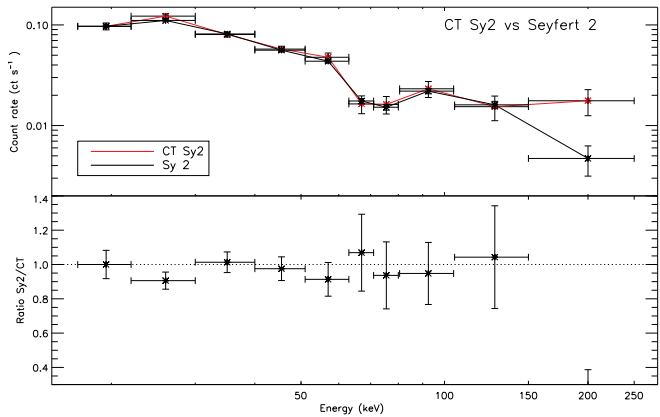}
\caption{Normalized spectra ({\it upper panel}) and ratio ({\it lower panel}) of Compton-thin Seyfert\,2s to CT Sy2s.}
\label{fig:sy2vsCT}
\end{figure}

In Fig.\,\ref{fig:sy1vs1.5}, we show the normalized spectra of Sy1s and Sy1.5s (in the upper panels) and their ratios (in the lower panel).
The average spectrum of Seyfert\,1s is consistent with that of Seyfert\,1.5s along the whole spectrum. 
The ratio of the average spectrum of Seyfert\,1 galaxies to the normalized one of NLS1s diverges from the unity from $\sim\,40$\,keV (Fig.\,\ref{fig:sy1vsNLS1}), with the spectrum of NLS1s being steeper than that of Sy1s.
Since the spectra of Seyfert\,1s and Seyfert\,1.5s are very similar, we merged them and compared the resulting spectrum to that of Seyfert\,2s. We show in Fig.\,\ref{fig:merged} the ratio obtained after normalization. The spectrum of Seyfert\,2s displays a clear excess over those of the Seyfert\,1s and Seyfert\,1.5s of $48\pm 8\%$ in the 22--63\,keV band. This bump peaks around 40\,keV, reaching 20\%, and might indicate that Seyfert\,2s have a stronger reflection component than Seyfert\,1s and Seyfert\,1.5s. At energies higher than 60\,keV, the ratio is consistent with 1, indicating similar cutoff energies.
The ratio of the spectra of Compton-thin to Compton-thick Seyfert\,2s is consistent with one up to 150 keV (Fig.\,\ref{fig:sy2vsCT}).
Including in our sample of Sy2s the two objects excluded because of their complex absorption does not affect the results, and the ratio between the spectra with and without these sources is consistent with one.

\subsection{Lightly obscured and mildly obscured Seyfert\,2s}\label{obs_unobs_modindep}

The Seyfert\,2s contained in our sample appear to have characteristics that differ significantly from those of Seyfert\,1 and Seyfert\,1.5 galaxies.
To test whether absorption might still play a role in the hard X-ray spectra of Compton-thin Seyfert\,2 galaxies, we divided the sample into two subsamples, one including the objects with $10^{23} \rm \,cm^{-2} \leq N_{\rm \,H}< 10^{24} \rm \,cm^{-2}$ (henceforth mildly obscured Seyfert\,2s, MOB\,Sy2s), and the other those with $N_{\rm \,H}< 10^{23} \rm \,cm^{-2}$ (henceforth lightly obscured Seyfert\,2s, LOB\,Sy2s). The MOB and LOB Sy2 samples contain 27 and 35 objects, respectively. In Table\,\ref{tab:flux}, we report the detection significances, exposures, and fluxes obtained for these samples. We did not consider the sources with an unknown value of $N_{\rm \,H}$. 

\begin{figure}[h]
\centering
\includegraphics[width=9cm]{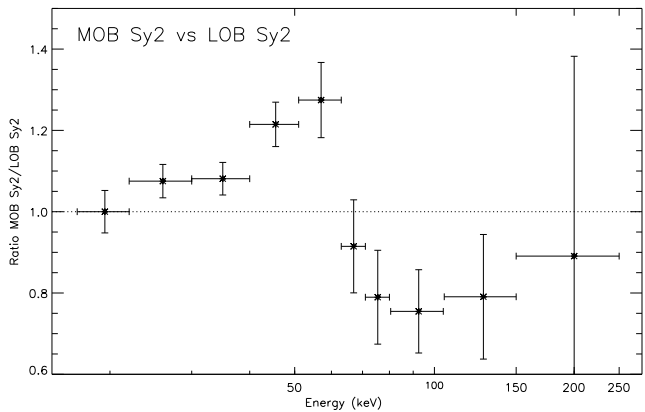}
\caption{Ratio of the normalized spectra of MOB to LOB Seyfert~2 galaxies.}
\label{fig:ratio_obs_unobs}
\end{figure}

\begin{figure}[h]
\centering
\includegraphics[width=9cm]{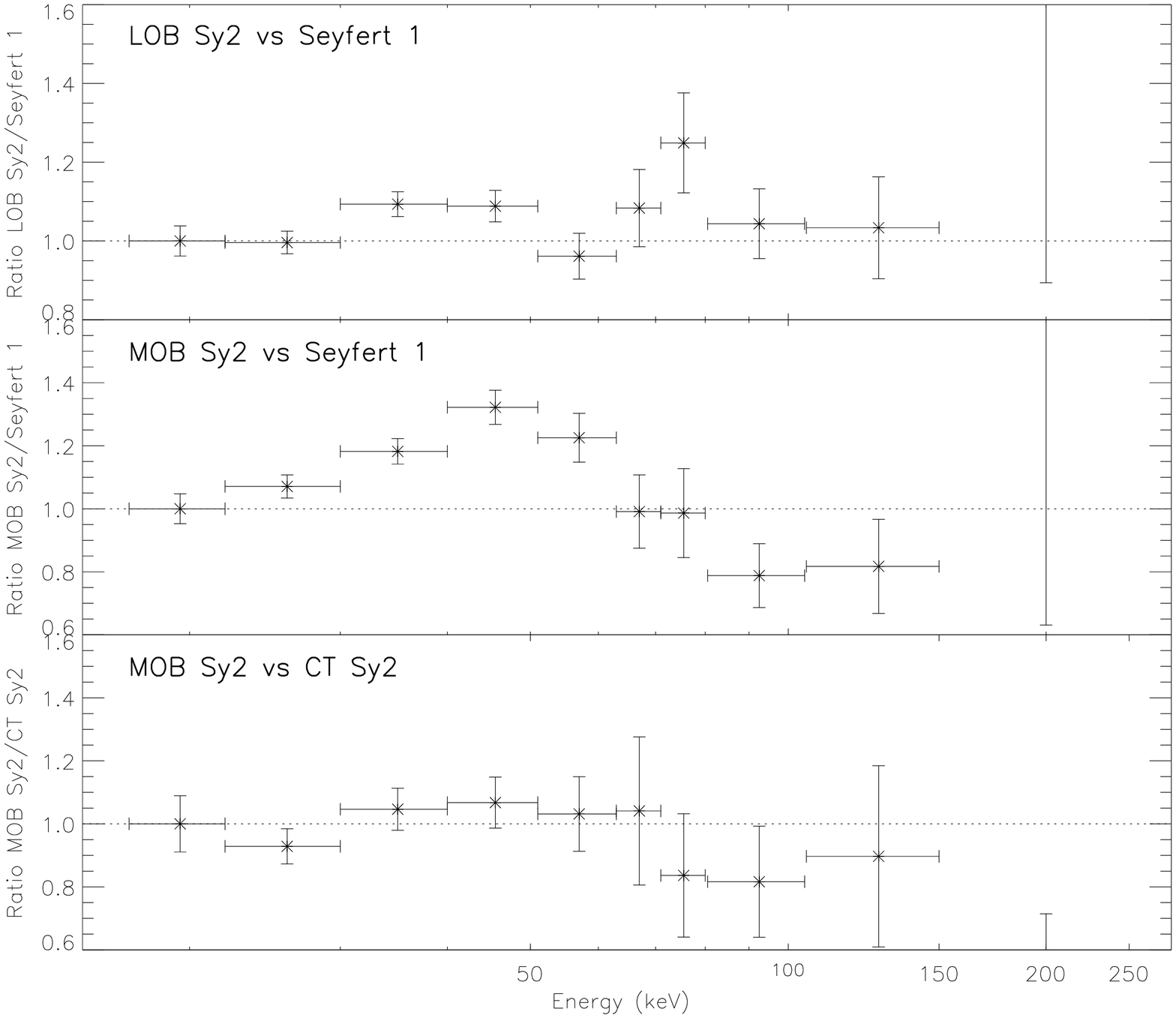}
\caption{Ratio between the normalized spectra of LOB Seyfert~2s and Seyfert~1s ({\it top panel}), between the spectra of MOB Seyfert~2s and Seyfert~1s ({\it center panel}), and between the spectra of MOB Seyfert~2s and CT Seyfert~2s ({\it bottom panel}).}
\label{fig:ratio_unobs_sy1}
\end{figure}

In Fig.\,\ref{fig:ratio_obs_unobs}, we show the ratio obtained by comparing the spectrum of MOB Sy2s to that of LOB Sy2s. The figure highlights significant differences between the two spectra, with the emission of MOB Sy2s being harder up to $\sim60$\,keV and then softer (up to $\sim\,150\,$keV) than that of LOB Sy2s. This clearly shows that the bump seen by comparing the spectrum of Seyfert\,2s to those of Seyfert\,1s and Seyfert\,1.5s (Fig.\,\ref{fig:merged}) comes prevalently from the contribution of MOB Sy2s (see Fig.\,\ref{fig:ratio_unobs_sy1}).

The spectrum of LOB Sy2s appears to be very similar to that of Seyfert\,1s (Fig.\,\ref{fig:ratio_unobs_sy1}), although an excess of $\simeq$ 20\% is evident in the 30--50 keV band. This excess might be related to a stronger reflection component in LOB Seyfert\,2s than in Seyfert\,1s. MOB Sy2s present instead a spectrum similar to that of CT Sy2s, and their ratio is consistent with 1 (Fig.\,\ref{fig:ratio_unobs_sy1}).

To test the influence of bright MOB Sy2s on their average spectrum, we compared the spectra obtained with and without the five brightest sources ($>20\sigma$). The ratio obtained after normalizing the two spectra is fully consistent with one, from which we can conclude that the average spectrum of MOB Sy2s is not influenced by peculiar bright sources.

\begin{figure}[]
\centering
\centering
\includegraphics[width=9cm]{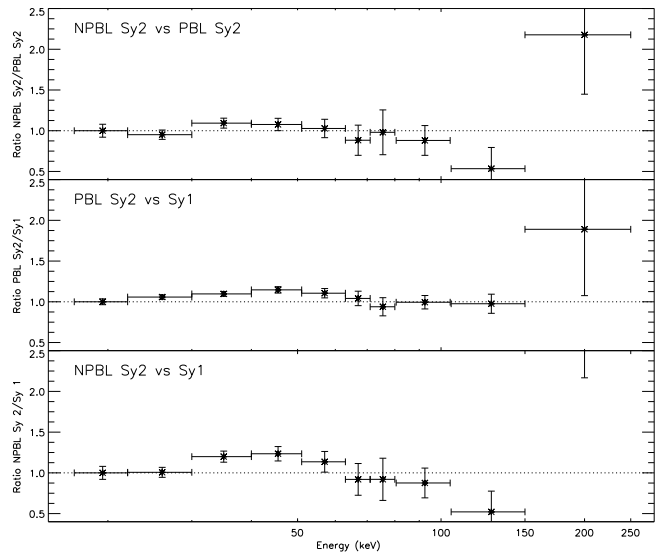}
  \caption{Ratio between the normalized spectra of PBL and NPBL Sy2s ({\it top panel}), between Seyfert~1s and PBL ({\it center panel}), and NPBL Sy2s ({\it bottom panel}).}
\label{fig:modIndependent_PBL}
\end{figure}

\subsection{Seyfert\,2s with and without PBLs}
To test whether PBL and NPBL Sy2s have different characteristics, we extracted and analyzed their average hard X-ray spectra. In Table\,\ref{tab:flux}, we report the detection significances, exposures, and fluxes obtained for the two samples.

Comparing the spectrum of PBL Sy2s to the one of NPBL Sy2s (Fig.\,\ref{fig:modIndependent_PBL}), it is evident that the ratio is consistent with 1.

When compared to the spectrum of Seyfert\,1s, the spectra of PBL and NPBL Sy2s have similar characteristics. As for the whole Seyfert\,2 sample, both PBL and NPBL Sy2s display evidence of a greater reflection component. This is probably due to the contribution of the MOB Sy2s present in the samples. 

\subsection{Comparison with a simulated power law spectrum}\label{sec:comppo}

\begin{figure}[h]
\centering
\includegraphics[width=9cm]{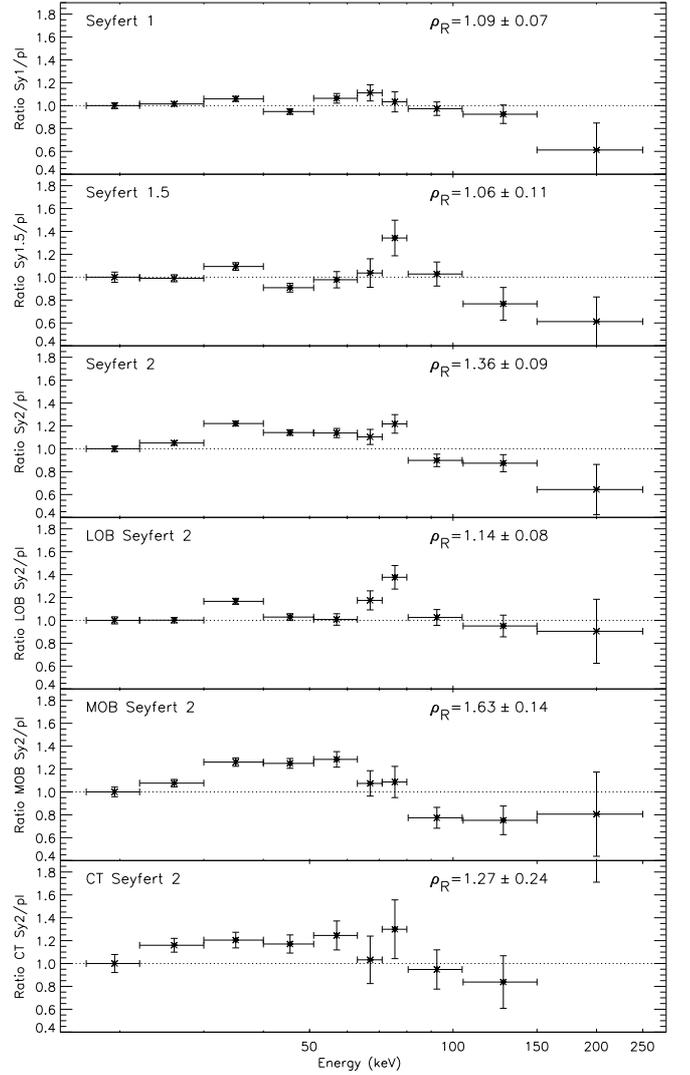}
\caption{Ratio between the normalized spectra of Seyfert~1s, Seyfert~1.5s, Compton-thin Seyfert~2, LOB Seyfert~2s, MOB Seyfert~2s, and Compton-thick Seyfert\,2s, and a simulated power law spectrum ($\Gamma=1.9$). The parameter $\rho_R$ is defined in Eq.\,\ref{rho}.}
\label{fig:syvsPo}
\end{figure}

In Fig.\,\ref{fig:syvsPo}, we show the ratios between the normalized spectra of Seyfert galaxies and a simulated power law spectrum with a photon index of $\Gamma=1.9$. 
Using XSPEC (version 12.5.0, Arnaud 1996), we simulated 20 power-law spectra, using the $fakeit$ command. We used the latest RMF and calculated the ARF from a weighted average of the nine available ARFs, with the weights proportional to the validity period duration of each ARF. The power law spectrum used is the average of the 20 simulated spectra, and the errors are their standard deviations. 

Seyfert\,1s and Seyfert\,1.5s show a ratio close to unity below 100 keV, and a small excess around 30 keV, which is possibly a signature of reflection. The excess in the 22-40 keV band is $8 \pm 3 \%$ and $9 \pm 3\%$ for Seyfert\,1s and Seyfert\,1.5s, respectively.

The spectrum of Seyfert\,2s displays a stronger excess ($27\pm 3 \%$) in the same band, and is harder than the power law up to 80 keV.
Above 150\,keV, the spectra of Seyfert\,1s, Seyfert\,1.5s and Seyfert\,2s become softer than the power law; this effect might be related to the presence of a high-energy cutoff.

LOB Sy2s show an excess of $17\pm 3\%$ over the simulated power law in the 22--40\,keV band, and do not present any significant softening above 100 keV.
MOB Sy2s show the most significant differences, being harder below $\simeq 60$ keV and softer above 80 keV. The excess is of $87\pm9\,\%$ in the 22--63\,keV band, and of $34\pm5\,\%$ in the 20--40 keV band. A similar excess ($78\pm 17 \%$ and $36\pm9\,\%$) is found by comparing the average spectrum of CT Seyfert\,2s to the simulated power law in the same bands. 

To estimate the amount of curvature in the spectrum, we introduce the parameter $\rho_R$ given by
\begin{equation}\label{rho}
\rho_R=\frac{R_{30}}{R_{80}},
\end{equation}
where $R_{30}$ and $R_{80}$ are the ratios of the normalized spectra to the power law in the 30--40 keV and 80--105 keV band, respectively. Seyfert\,1s ($\rho_R= 1.09\pm 0.07$) have a value of $\rho_R$ consistent with those of Seyfert\,1.5s ($\rho_R= 1.06\pm 0.11$) and LOB Sy2s ($\rho_R= 1.14\pm 0.03$). The value of this parameter is significantly larger for MOB Sy2s ($\rho_R= 1.63\pm0.14$), while it is $\rho_R= 1.27\pm 0.24$ for CT Sy2s.

\section{Model-dependent spectral analysis}\label{spectral_analysis}
We analyzed the average IBIS/ISGRI spectra using XSPEC. 
Throughout the paper, we used 3$\sigma$ errors. The probabilities that additive parameters improve a fit are calculated with the F-test. 

We tested simple models such as a power law, and a power law with an exponential cutoff at high energies.
To account for the presence of reflection, we used the {\it pexrav} model (Magdziarz \& Zdziarski 1995). {\it Pexrav} calculates the spectrum produced by an X-ray source reflected by an optically thick neutral slab. In this model, the parameter $R$ measures the reflection component. If the source is isotropic, $R$ can be linked to the solid angle $\Omega$ as $R\simeq \frac{\Omega}{2 \pi}$.
The value of $R$ depends on the inclination angle $i$ between the perpendicular to the accretion disk and the line of sight. The smaller the inclination angle, the larger the resulting reflection component. The quality of the data does not allow us to constrain both $R$ and $i$ at the same time, so we fixed the inclination angles to $i=45^{\circ}$ for all the classes. We used the same value of $i$ for all the classes to characterize the reflection with a single parameter. We note that by doing this $R$ no longer represents $\Omega/2\pi$, and becomes only an indicator of the reflection amplitude.

\subsection{Seyfert\,galaxies}
\label{Sy_spanalysis}

Fitting the spectra of Seyfert\,1s, Seyfert\,1.5s, and Seyfert\,2s with a simple power law yields a reduced chi-squared of $\chi^{2}_{\nu}=1.4$, $\chi^{2}_{\nu}\simeq 1.4$, and $\chi^{2}_{\nu}\simeq 3.9$, respectively. The null-hypothesis probability is 20\%, 18\%, and $<0.1\%$ for Sy1s, Sy1.5s, and Sy2s, respectively.

Applying a cutoff power law improves the fit of both Seyfert\,1s and Seyfert\,2s at a confidence level of 84\% and 99.6\%, respectively. Adding a reflection component to the power law model also significantly improves the fit for both Seyfert\,1s and Seyfert\,2s. The fit to the spectrum of Seyfert\,1.5s does not improve significantly by adding any of these features to the baseline power law model. Fitting with different models the spectrum of Seyfert\,1.5s, one notes that most of the chi-squared comes from a single bin (71--80 keV, see Appendix\,\ref{Sec:spectraIm}). On the basis of this, and that in Sect.\,\ref{sect:modIndep} we have shown that the spectra of Seyfert\,1s and Seyfert\,1.5s are very similar, we use the same models to characterize their spectra. The results of the spectral analysis are reported in Table\,\ref{tab:Sy_an}.

Both a high-energy cutoff and a reflection component from neutral matter are thought to play an important role in the hard X-ray emission of Seyfert galaxies, thus in the following we discuss only the results obtained using $pexrav$.
Applying this model, we obtained a lower limit to the high-energy cutoff of $E_C\gtrsim 190$\,keV, and a reflection normalization of $R\leq 0.4 $ for both Seyfert\,1s and Seyfert\,1.5s. The consistency of these parameters agrees with that found in Sect.\,\ref{sect:modIndep}. Using the same model for the average spectrum of Seyfert\,2s, we obtained $E_{C}=154^{+185}_{-84}$ keV and $R \leq 2.6$, which confirms the greater curvature observed.

Owing to its low statistics, a simple power law model provides a good fit ($\chi^{2}_{\nu}= 1$) to the stacked spectrum of the ten CT Seyfert\,2 galaxies of our sample.  Although statistically not required, a cutoff and a reflection component reduce the chi-squared of $\Delta \chi^{2}=3.8$ and $\Delta \chi^{2}=5.1$, respectively.
Using $pexrav$, we could not constrain $E_C$, while we obtained an upper limit to the reflection normalization of $R\leq11.2$.

A power-law model cannot represent well the hard X-ray spectrum of NLS1s. Adding a cutoff or a reflection component significantly improves the fit, although it is not possible to constrain the two parameters at the same time. We obtained a photon index of $\Gamma=2.3^{+0.1}_{-0.6}$, a lower limit to the energy of the cutoff of $E_{C}\geq 53$\,keV, and to the reflection parameter of $R\geq 0.1$.

The average photon indices obtained are consistent for the different subsamples of Seyfert galaxies, and have values of $\Gamma\simeq 1.8$, in agreement with previous studies (e.g., Dadina et al. 2008). The spectra of Seyfert\,1s, Seyfert\,1.5s, Seyfert\,2s, CT Seyfert\,2s, and NLS1s are shown in Appendix\,\ref{Sec:spectraIm}.

\subsection{MOB and LOB Seyfert\,2s}
\label{ObsUnobs_spanalysis}
Using $pexrav$ to fit the spectra of MOB and LOB Sy2s, we confirmed the different curvatures of the spectra of the two subsamples, obtaining values of the reflection parameter of $R\leq 0.5$ and $R=2.2^{+4.5}_{-1.1}$, for LOB and MOB Sy2s, respectively. The photon indices ($\Gamma\sim 1.8$) and the energies of the cutoff ($E_C^{\tiny \rm\,LOB}=425^{+267}_{-120}$\,keV and $E_C^{\tiny \rm\,MOB}=287^{+105}_{-64}$\,keV) obtained by the fit are consistent to within 3$\sigma$ . The results are reported in Table\,\ref{tab:Sy_an_mob}, while the spectra of MOB and LOB Sy2s are shown in Appendix\,\ref{Sec:spectraIm}.

The spectral analysis performed using different models confirms that the average hard X-ray spectrum of LOB Sy2s is more similar to those of Seyfert\,1s and Seyfert\,1.5s than to that of MOB Sy2s.

\subsection{Seyfert\,2s with and without PBL}
\label{spec_PBL}

Consistently with what was obtained for the whole sample of Seyfert\,2s, the spectra of PBL and NPBL Sy2s cannot be reproduced using solely a power law, but need to be fitted using a more complex model. 
Adding a high-energy cutoff or a reflection component to the power law significantly improves the fit for both spectra. All the models used give values of the parameters of the two spectra that are consistent to within $3 \sigma$, confirming the similarities found by the model-independent analysis. The results of the spectral analysis are reported in Table\,\ref{tab:Sy_an_pbl}.

\begin{table}[t]
\centering
\caption{Results obtained from the spectral analysis of the average hard X-ray spectra of Seyfert\,1s, Seyfert\,1.5s, Seyfert\,2s, CT Sy2s, and NLS1s. The inclination angle $i$ was fixed to $45^{\circ}$.}
\resizebox{\columnwidth}{!}{
\label{tab:Sy_an}
\begin{tabular}[c]{lccccc}
\hline \hline 
\noalign{\smallskip}
Model &$\Gamma$&$E_{C}$& R &$\chi ^2$/DOF& FTEST\\
               & & [{\tiny keV}] &  & & \\
\noalign{\smallskip\hrule\smallskip}
\multicolumn{6}{l}{{\bf Seyfert\,1s}} \\
\noalign{\smallskip\hrule\smallskip}
Power law      &$ 1.97  ^{+0.05}_{-0.05}$&  -- & --  &  11/8 & -- \\
\noalign{\smallskip}
Cut-off power law       &$ 1.8  ^{+0.1}_{-0.2}$& $ 293^{+NC}_{-150} $ & --  & 8.2/7 & 0.17 \\
\noalign{\smallskip}
$^*$Pexrav         &$ 1. 96^{+0.04}_{-0.04}  $& --  & $ 0.2^{+0.2}_{-0.2}  $ & 9.8/7 & 0.38  \\      
\noalign{\smallskip}
Pexrav         &$ 1.8 ^{+0.2}_{-0.1}  $&$ 340^{+NC}_{-155} $ & $ 0.1^{+0.3}_{-0.1} $ & 8.3/6 & 0.43 \\      
\noalign{\smallskip\hrule\smallskip}
\multicolumn{6}{l}{{\bf Seyfert\,1.5s}} \\
\noalign{\smallskip\hrule\smallskip}
Power law      &$  1.97  ^{+0.09}_{-0.08}$& -- & --  & 11.4/8 &  -- \\
\noalign{\smallskip}
Cut-off power law       &$ 1.9  ^{+0.1}_{-0.3} $& $ 500^{+310}_{-250} $ &--  & 11.3/7 & 0.81  \\
\noalign{\smallskip}
$^*$Pexrav         &$ 1. 97^{+0.05}_{-0.05}  $& -- & $ 0.2^{+0.3}_{-0.2}  $ & 11.3/7 & 0.81 \\     
\noalign{\smallskip}
Pexrav         &$ 1.8  ^{+0.2}_{-0.1}  $&$332^{+NC}_{-145} $  & $ 0.1^{+0.3}_{-0.1} $ & 11.3/6 & 0.97 \\      
\noalign{\smallskip\hrule\smallskip}
\multicolumn{6}{l}{{\bf Seyfert\,2s}} \\
\noalign{\smallskip\hrule\smallskip}
Power law      &$ 1.93^{+0.03}_{-0.03} $& -- & --  &  31/8 & -- \\
\noalign{\smallskip}
Cut-off power law       &$  1.4 ^{+0.2}_{-0.2} $& $86 ^{+41}_{-22} $  & --  & 8.9/7  & $<0.01$ \\
\noalign{\smallskip}
$^*$Pexrav         &$ 1. 97^{+0.05}_{-0.05}  $& --   & $ 2.0^{+3.9}_{-1.2}  $ & 9.3/7 & $<0.01$ \\     
\noalign{\smallskip}
Pexrav         &$  1.6  ^{+0.4}_{-0.4}   $&$ 154 ^{+185}_{-84} $  & $   0.4^{+2.2}_{-0.4} $ & 7.4/6 & 0.01  \\      
\noalign{\smallskip\hrule\smallskip}
\multicolumn{6}{l}{{\bf CT Seyfert\,2s}} \\
\noalign{\smallskip\hrule\smallskip}
Power law      &$ 1.9^{+0.1}_{-0.1}  $& -- & -- &  8/8  & -- \\
\noalign{\smallskip}
Cut-off power law      &$  1.5 ^{+0.2}_{-0.2}   $& $ 100^{+NC}_{-51}  $  & --  & 4.2/7 & 0.04  \\
\noalign{\smallskip}
$^*$Pexrav        &$  1.9  ^{+0.1}_{-0.1}   $& --  & $ 1.4 ^{+9.6}_{-1.2}   $ &  2.7/7 & $<0.01$ \\    
\noalign{\smallskip}
Pexrav        &$  2.0  ^{+0.1}_{-0.1}   $&$  NC  $  & $ 1.5 ^{+9.7}_{-1.5}  $ &  2.7/6 & 0.04 \\    

\noalign{\smallskip\hrule\smallskip}
\multicolumn{6}{l}{{\bf Narrow Line Seyfert\,1s}} \\
\noalign{\smallskip\hrule\smallskip}
Power law      &$ 2.23  ^{+0.08}_{-0.08} $& -- & --  & 24.6/8  & -- \\
\noalign{\smallskip}
Cut-off power law       &$ 1.7  ^{+0.2}_{-0.6}  $& $ 70 ^{+43}_{-36} $ &--  & 11.6/7 & 0.03 \\
\noalign{\smallskip}
$^*$Pexrav         &$  2.28 ^{+0.08}_{-0.08}  $& -- & $  4.3^{+NC}_{-3.0} $ & 9.5/7 & 0.01 \\     
\noalign{\smallskip}
Pexrav         &$  2.3 ^{+0.1}_{-0.6}  $&$ 310 ^{+NC}_{-257} $ & $  4.2 ^{+NC}_{-4.1} $ & 9.5/6 & 0.06 \\     
\noalign{\smallskip\hrule\smallskip}
\noalign{\smallskip}
\multicolumn{6}{l}{{\bf Notes.} {\footnotesize $NC$: the parameter or the 3$\sigma$ error were not constrained.  }  } \\
\multicolumn{6}{l}{{\footnotesize $^*$Pexrav: the high-energy cut off was not included, being fixed to } } \\
\multicolumn{6}{l}{{\footnotesize  its upper limit $E_C=10^6$ keV. } } \\
\end{tabular}
}
\end{table}

\begin{table}[t]
\centering
\caption{Results obtained from the spectral analysis of the average hard X-ray spectra of MOB and LOB Seyfert\,2s. The inclination angle $i$ was fixed to $45^{\circ}$.}
\resizebox{\columnwidth}{!}{
\label{tab:Sy_an_mob}
\begin{tabular}[c]{lccccc}
\hline \hline \noalign{\smallskip}
Model &$\Gamma$&$E_{C}$&  R &$\chi ^2$/DOF& FTEST\\
               & & [{\tiny keV}] &   & &\\
\noalign{\smallskip}\hline\noalign{\smallskip}
\multicolumn{6}{l}{{\bf MOB Seyfert\,2s}} \\
\noalign{\smallskip\hrule\smallskip}
Power law     &$  1.94^{+0.05}_{-0.05}  $ & --   & --  & 45.2/8 & -- \\
\noalign{\smallskip}
Cut-off power law       &$  1.0^{+0.3}_{-0.3}   $& $ 47^{+20}_{-11}  $ & --   & 7.6/7 & $<0.01$ \\
\noalign{\smallskip}
$^*$Pexrav      &$   1.95^{+0.05}_{-0.05}   $& -- &  $  7^{+7}_{-2}  $ & 9.8/7 & $<0.01$ \\    
\noalign{\smallskip}
Pexrav         &$  1.82^{+0.06}_{-0.06}  $&$  287^{+105}_{-64} $  & $  2.2^{+4.5}_{-1.1}  $ &  8.8/6 & $<0.01$ \\    
\noalign{\smallskip\hrule\smallskip}
\multicolumn{6}{l}{{\bf LOB Seyfert\,2s}} \\
\noalign{\smallskip\hrule\smallskip}
Power law     &$  1.91^{+0.04}_{-0.04} $  & -- & --  & 15.9/8 & --  \\
\noalign{\smallskip}
Cut-off power law       &$  1.7^{+0.2}_{-0.1}   $& $ 224^{+NC}_{-98}  $ & --   &  12.8/7 & 0.23 \\
\noalign{\smallskip}
$^*$Pexrav      &$   1.91^{+0.04}_{-0.04}   $& --  &  $ 0.4^{+0.5}_{-0.3} $ & 13.7/7 & 0.32 \\    
\noalign{\smallskip}
Pexrav         &$   1.8 ^{+0.2}_{-0.3}  $&$ 425^{+267}_{-120}  $ &  $ 0.2^{+0.3}_{-0.2} $ &  12.9/6 & 0.53 \\    
%
%
%
%
%
\noalign{\smallskip}
\hline
\noalign{\smallskip}
\multicolumn{6}{l}{{\bf Notes.} {\footnotesize $NC$: the parameter or the 3$\sigma$ error were not constrained.  }  } \\
\multicolumn{6}{l}{{\footnotesize $^*$Pexrav: the high-energy cut off was not included, being fixed to } } \\
\multicolumn{6}{l}{{\footnotesize  its upper limit $E_C=10^6$ keV. } } \\
\end{tabular}
}
\end{table}

\begin{table}[t]
\centering
\caption{Results obtained from the spectral analysis of the average hard X-ray spectra of PBL and NPBL Seyfert\,2s. The inclination angle $i$ was fixed to $45^{\circ}$.}
\resizebox{\columnwidth}{!}{
\label{tab:Sy_an_pbl}
\begin{tabular}[c]{lccccc}
\hline \hline \noalign{\smallskip}
Model &$\Gamma$&$E_{C}$& R &$\chi ^2$/DOF& FTEST\\
               & & [{\tiny keV}] &  &&\\
\noalign{\smallskip\hrule\smallskip}
\multicolumn{6}{l}{{\bf PBL Seyfert\,2s}} \\
\noalign{\smallskip\hrule\smallskip}
Power law     &$   1.94^{+0.03}_{-0.03}  $ & --  & --  & 23.7/8 &  -- \\
\noalign{\smallskip}
Cut-off power law       &$  1.6^{+0.2}_{-0.2}   $& $ 136^{+135}_{-48}  $  & -- & 11.7/7 & 0.03 \\
\noalign{\smallskip}
%
%
$^*$Pexrav      &$   1.95 ^{+0.03}_{-0.03}   $& --    &  $  1.0^{+0.6}_{-0.4}   $ & 8.5/7 & $<0.01$ \\    
\noalign{\smallskip}
Pexrav         &$   1.9^{+0.2}_{-0.3}   $&$  803^{+NC}_{-660} $  & $  0.7^{+0.8}_{-0.6}  $ & 8.5/6 & 0.05  \\    

\noalign{\smallskip\hrule\smallskip}
\multicolumn{6}{l}{{\bf NPBL Seyfert\,2s}} \\
\noalign{\smallskip\hrule\smallskip}
Power law     &$   1.95^{+0.09}_{-0.09}  $ & -- & -- &  22.9/8 & -- \\
\noalign{\smallskip}
Cut-off power law       &$  1.1^{+0.5}_{-0.5}  $& $ 49^{+70}_{-20} $ & --  &  13.9/7 & 0.07 \\
\noalign{\smallskip}
%
%
$^*$Pexrav      &$    1.97^{+0.09}_{-0.09}   $& --   &  $ 3^{+NC}_{-2.3}  $ &  14.4/7 & 0.08 \\    
\noalign{\smallskip}
Pexrav         &$    1.9^{+0.1}_{-0.1}   $&$  874^{+NC}_{-548}  $  & $  2.9^{+NC}_{-2.3}  $ &  13.7/6 & 0.21 \\    
\noalign{\smallskip}
\hline
\noalign{\smallskip}
\multicolumn{6}{l}{{\bf Notes.} {\footnotesize $NC$: the parameter or the 3$\sigma$ error were not constrained.  }  } \\
\multicolumn{6}{l}{{\footnotesize $^*$Pexrav: the high-energy cut off was not included, being fixed to } } \\
\multicolumn{6}{l}{{\footnotesize  its upper limit $E_C=10^6$ keV. } } \\
\end{tabular}
}
\end{table}

\begin{figure*}[!t]
\centering
\includegraphics[width=9.15cm]{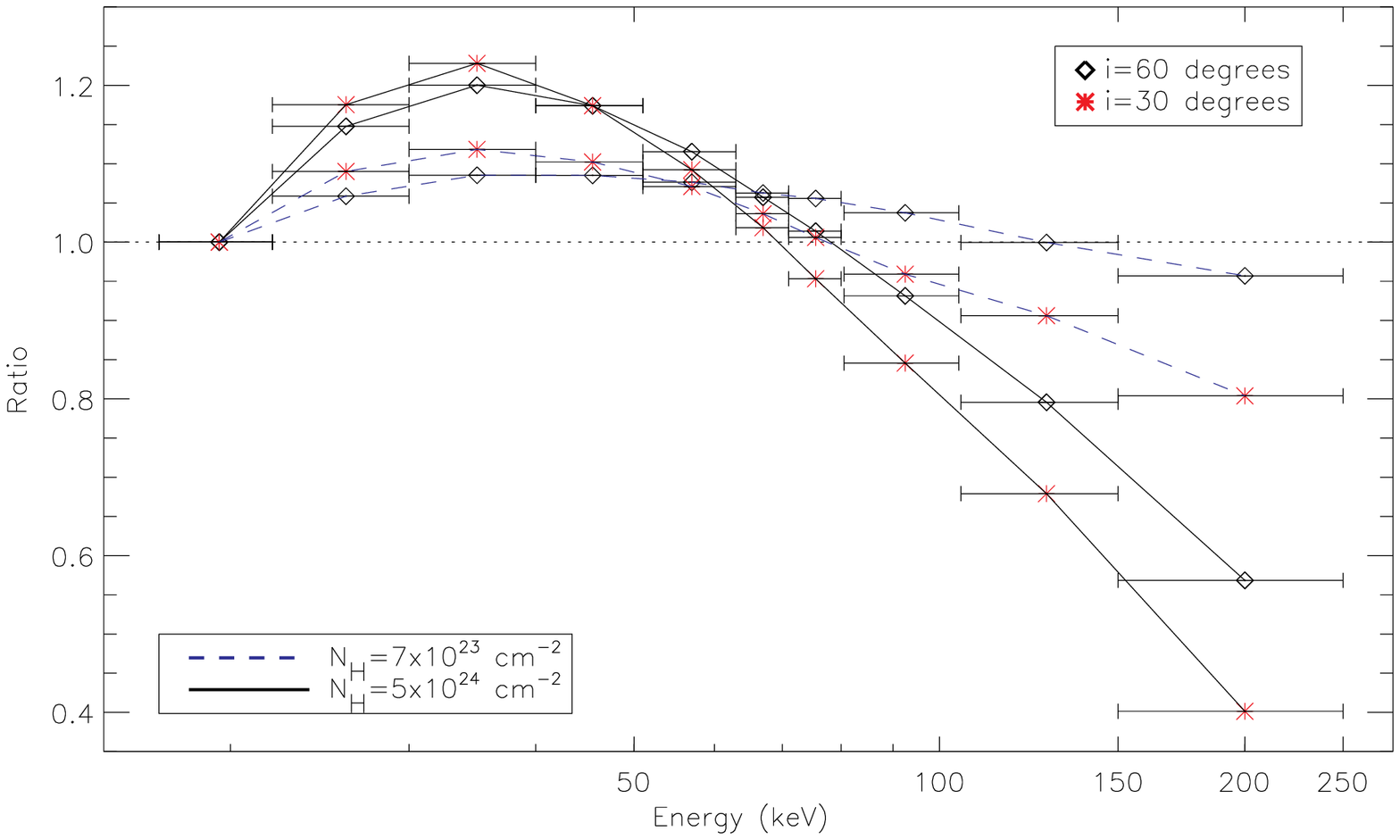}
\includegraphics[width=9.15cm]{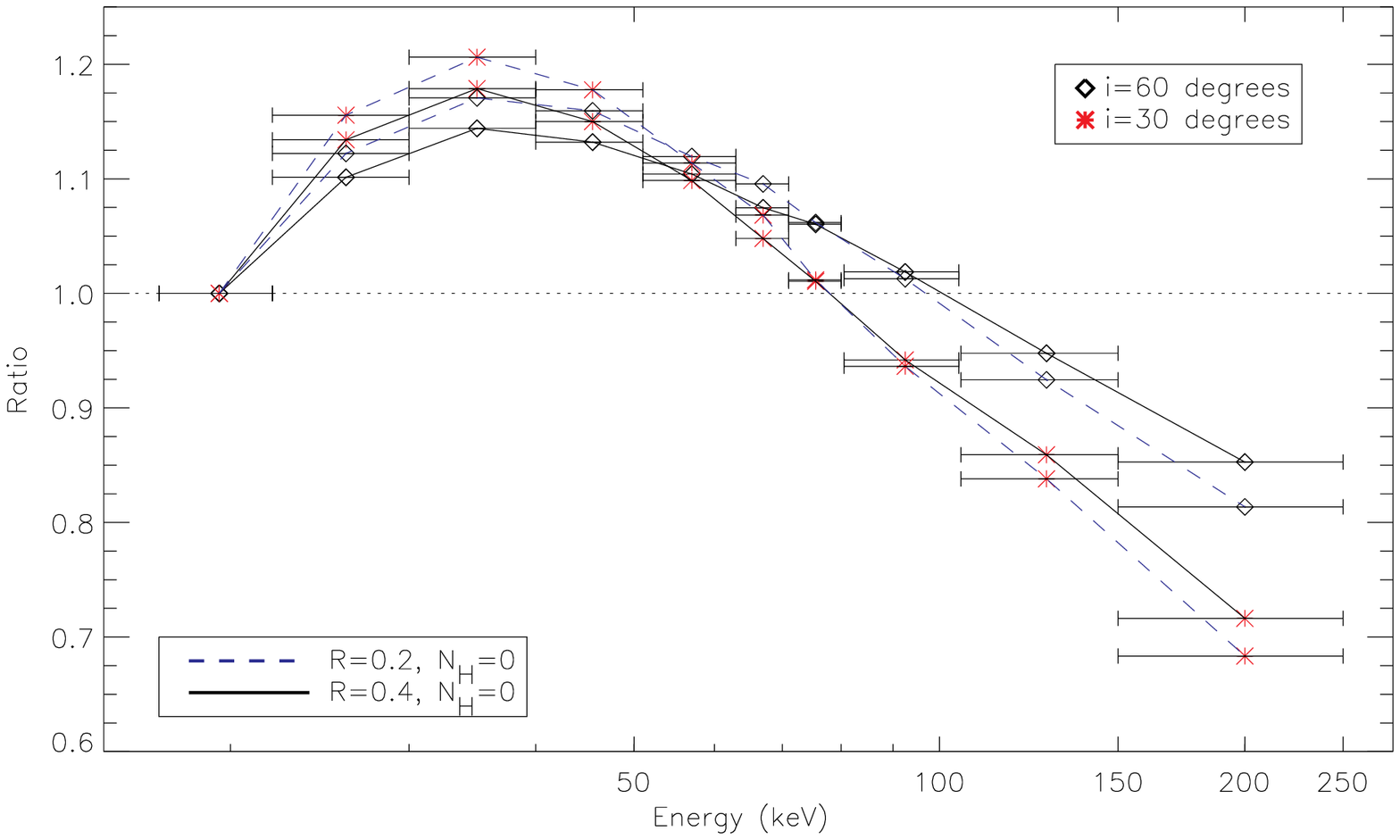}
 \caption{Ratios of simulated spectra of MOB (Eq.\,\ref{eq:sim1i}) to LOB (Eq.\,\ref{eq:sim1a}) objects. We assumed $\Gamma=1.8$, $E_C=300$\,keV for the continuum of all spectra. We tested different values of $i$ and $N_{\rm \,H}$ ($N_{\rm \,H}=7\times10^{23}\rm\,cm^{-2}$ and $N_{\rm \,H}=5\times10^{24}\rm\,cm^{-2}$) for the spectrum of MOB Sy2s, assuming $R=1$ for all the spectra, and $N_{\rm \,H}=0$ for LOB Sy2s ({\it left panel}). We tested different values of the reflection parameter ($R=0.4$ and $R=0.2$ for LOB objects, fixing $R=1$ for MOB Sy2s), and of $i$ (only for MOB Sy2s), setting $N_{\rm \,H}=3\times10^{23}\rm\,cm^{-2}$ for MOB Sy2s and $N_{\rm \,H}=0$ for LOB Sy2s ({\it right panel}).
 }
\label{fig:Refl_sim}
\end{figure*}

\section{Reflection in MOB Sy2s}\label{sect:bump}
We have shown that the spectrum of MOB Sy2s shows a stronger reflection component than those of LOB Sy2s, Seyfert\,1s, and Seyfert\,1.5s, and we now investigate the possible causes of this effect. 

The spectrum of lightly obscured objects (LOB Sy2s, Sy1s, and Sy1.5s) can be expressed by
\begin{equation}\label{eq:sim1a}
F(E)=AE^{-\Gamma}e^{-E/E_C} + R(E),
\end{equation}
where $A$ is the normalization and $R(E)$ the reflection component, which was accounted for using $pexrav$. 
The observed 20--60 keV bump might be due to a distant (i.e. unabsorbed) reflection component, which becomes more important for MOB Sy2s due to the larger fraction of Compton scattered continuum. This reflector has been associated with the inner wall of the torus, or the outer part of the disk. The spectrum of MOB objects can be represented by
\begin{equation}\label{eq:sim1i}
F(E)=M(E,N_{\rm \,H})AE^{-\Gamma}e^{-E/E_C} + R(E).
\end{equation}
In the following, we assume that $\Gamma=1.8$, $E_C=300$\,keV.
We first simulated and compared several spectra with the same continuum and reflection component, but with different values of $N_{\rm \,H}$, to see whether absorption might explain the differences observed. The value of R associated with the torus remains unknown (e.g., Gilli et al. 2007 consider a value of $R_{\rm \,T}=0.37$). To obtain an upper limit to the effect, we assumed $R=1$ (for both LOB and MOB objects), and column densities of $N_{\rm \,H}=7\times 10^{23}\rm \,cm^{-2}$ and $N_{\rm \,H}=0$ for MOB and LOB objects, respectively. We tested two different values of the inclination angle for MOB Sy2s ($i=30^{\circ}$ and $i=60^{\circ}$), while we set $i=60^{\circ}$ for LOB objects. The parameter $M(E,N_{\rm \,H})$ includes both Compton scattering (through the $cabs$ model) and photoelectric absorption (implemented as $wabs$), as given in Eq.\ref{eq:abs}. In Fig.\,\ref{fig:Refl_sim} ({\it left panel}), we show the ratios obtained comparing the simulated spectra. From the figure, it is evident that a column density of $N_{\rm \,H}=7\times 10^{23}\rm \,cm^{-2}$ is largely insufficient to explain the observed differences. To provide an excess closer to what is observed, one would need a value of the column density as large as $N_{\rm \,H}=5\times 10^{24}\rm \,cm^{-2}$.

We then tested different values of the reflection parameter. From the spectral analysis, we found upper limits of the reflection parameter of $R\leq 0.4$ for Seyfert\,1s and Seyfert\,1.5s, and of $R\leq 0.5$ for LOB\,Sy2s. These values are significantly lower than that obtained for MOB Sy2s ($R=2.2^{+4.5}_{-1.1}$). We simulated and compared spectra obtained using Eqs.\,\ref{eq:sim1a} and \ref{eq:sim1i}, considering different amounts of reflection and absorption for LOB ($R=0.5$ and $R=0.2$, $N_{\rm \,H}=0$, $i=60^\circ$) and MOB ($R=1$, $N_{\rm \,H}=7\times 10^{23}\rm \,cm^{-2}$, $i=60^\circ$ and $i=30^\circ$) objects. From Fig.\,\ref{fig:Refl_sim} ({\it right panel}), one can see that a greater unabsorbed reflection component in MOB Seyfert\,2s would explain their larger amount of  flux in the 20--60 keV energy band with respect to LOB objects.

We fitted the spectrum of MOB Sy2s with Eq.\,\ref{eq:sim1i} in XSPEC, setting the column density of the absorbing medium to the average value (weighted by the detection significance) of the sample ($N_{\rm \,H}=3\times10^{23}\rm \,cm^{-2}$). 
We first assumed a edge-on reflector ($i=60^{\circ}$), and fixed $E_C$ to different values. The best fit yields $R\geq 2$ and is obtained for $E_C=300$ keV. The value of the reflection parameter is $R>1$ even when considering higher values of $E_C$.
Considering a face-on reflector ($i=30^{\circ}$), we obtained a lower value of the reflection parameter for $E_C=300$\,keV ($R=1.1^{+3.5}_{-0.5}$), while we obtained $R\geq 1$ for higher values of $E_C$. The results of the fits are listed in Table\,\ref{tab:obscured_xspec}.

\begin{table}[t]
\centering
\caption{Results obtained by fitting the spectrum of MOB Seyfert\,2s with Eq.\,\ref{eq:sim1i}, fixing $N_{\rm \,H}$ to the average value of the sample. We tested different values of the cutoff energy and the inclination angle.}
\label{tab:obscured_xspec}
\begin{tabular}[c]{cccccc}
\hline \hline \noalign{\smallskip}
$E_C$&$\Gamma$&R& $i$ &$\chi ^2$&DOF\\
   {\tiny [keV]}  &   &   & & & \\
\noalign{\smallskip\hrule\smallskip}
\noalign{\smallskip}
300$^*$   &$  1.78^{+0.07}_{-0.07}  $ & $3^{+10}_{-1}$ & $ 60^{\circ}$  &  9.6 & 7  \\
\noalign{\smallskip}
       &$   1.74^{+0.06}_{-0.06}  $ & $ 2^{+3 }_{-1}$ & $ 45^{\circ}$  &  9.7  & 7  \\
\noalign{\smallskip}
       &$  1.76^{+0.07}_{-0.09}  $ & $1.3^{+1.8}_{-0.5}$ & $ 30^{\circ}$  &   9.9 & 7  \\
\noalign{\smallskip}
500$^*$   &$  1.83^{+0.07}_{-0.07}  $ & $3.5^{+NC}_{-2.0}$ & $ 60^{\circ}$  & 10.3  &  7  \\
\noalign{\smallskip}
       &$  1.82^{+0.06}_{-0.06}  $ & $2.5^{+ 7.8}_{-1.5}$ & $ 45^{\circ}$  & 10.3 & 7  \\
\noalign{\smallskip}
   &$  1.81^{+0.07}_{-0.07}  $ & $2.0^{+3.5}_{-1}$ & $ 30^{\circ}$  & 10.4  &  7  \\
\noalign{\smallskip}
$\infty$$^*$  &$  1.99^{+0.06}_{-0.06}  $ & $13^{+NC}_{-10}$ & $ 60^{\circ}$  & 11.9  &  7  \\
\noalign{\smallskip}
       &$  1.95^{+0.06}_{-0.05}  $ & $5.3^{+NC}_{-3.7}$ & $ 45^{\circ}$  & 11.8  & 7  \\
\noalign{\smallskip}
  &$  1.96^{+0.07}_{-0.07}  $ & $3^{+NC}_{-2}$ & $ 30^{\circ}$  & 11.9  &  7  \\
\noalign{\smallskip}
\hline
\noalign{\smallskip}
\multicolumn{6}{l}{{\bf Notes.} {\footnotesize $NC$: 3$\sigma$ error not constrained; $^*$: parameter fixed.}} \\
\end{tabular}
\end{table}

\section{Discussion}\label{sect:discussion}
We now discuss the average hard X-ray spectra of the different classes of Seyfert galaxies, the differences and similarities between PBL and NPBL Sy2s, the average hard X-ray emission of NLS1s, and all this in the framework of the unified model and the cosmic X-ray background.

\subsection{Seyfert galaxies: the continuum}
From our analysis of the average hard X-ray spectra of Seyfert galaxies, we found that the primary power law emission is similar for both Seyfert\,1s and Seyfert\,2s, and can be described by a photon index of $\Gamma \simeq 1.8$.
Comparing the average spectra to a simulated power law, we found for Sy1s, Sy1.5s and Sy2s only a weak softening in the last energy bin, which might imply that on average the cutoff is at energies $E_{C}\gtrsim150$\,keV. This was confirmed by the model-dependent spectral analysis, from which we found $E_C\gtrsim 200$\,keV, which is consistent with the results of {\it BeppoSAX} observations (Dadina 2008).

In the thermal Comptonization scenario, the power law emission of AGN is commonly assumed to arise from inverse Comptonization of soft photons in a hot plasma. Within this scenario, the cutoff energy and the photon index are linked to the temperature and the optical depth of the hot plasma responsible for the Comptonization. According to Petrucci et al. (2001), the temperature can be estimated as $kT_e=E_C/2$ for $\tau \lesssim 1$, or $kT_e=E_C/3$ for $\tau \gg 1$, and the three parameters are related by
\begin{equation}\label{eq:Petrucci}
\Gamma-1\simeq \left\{\frac{9}{4}+\frac{m_ec^2}{kT_e\tau (1+\tau/3)}\right\}^{1/2}-\frac{3}{2}.
\end{equation}
Using Eq.\,\ref{eq:Petrucci} and the values of the parameters obtained using the {\it pexrav} model, we calculated the average optical depths of the different classes of Seyferts. For Seyfert\,1s and Seyfert\,1.5s, we obtained $\tau=0.8^{+0.6}_{-0.5}$, which is consistent with the values obtained for LOB ($\tau=0.7^{+0.6}_{-0.4}$) and MOB Seyfert\,2s ($\tau_{2}=0.9^{+0.3}_{-0.3}$). The consistency of $\tau$ indicates that the average physical characteristics of the Comptonizing medium are similar for different classes of Seyferts, as expected from the UM.

\subsection{Seyfert galaxies: the reflection component}
The hard X-ray spectrum of Seyfert\,2s shows a harder emission in the 20--60\,keV band than those of Seyfert\,1s and Seyfert\,1.5s (Fig.\,\ref{fig:merged}). 
Most of this difference can be ascribed to the contribution of MOB Sy2s (Fig.\,\ref{fig:ratio_obs_unobs}), while LOB Sy2s show characteristics more similar to Seyfert\,1s and Seyfert\,1.5s (Fig.\,\ref{fig:ratio_unobs_sy1}), although still having a $\sim 20\%$ excess in the 30--50 keV band. 

The greater reflection of MOB Sy2s cannot be explained solely by the dampening of the continuum caused by absorption, but we have demonstrated that different values of $R$ must also be taken into account (Fig.\,\ref{fig:Refl_sim}). We have in fact shown that the average value of Seyfert\,1/1.5s and LOB Sy2s is $R\leq 0.4$ and $R\leq 0.5$ (Section\,\ref{spectral_analysis}), respectively, while for MOB Sy2s this value is greater ($R>1$, see Section\,\ref{sect:bump}).

From a purely geometrical point of view, and considering the disk as the main reflector, one would expect a greater influence of the reflection component in the spectrum of Seyfert\,1s rather than in that of Seyfert\,2s, the angle $i$ between the normal to the disk and the observer being smaller. 
Different luminosity distributions could introduce a bias in our hard X-ray selected sample of MOB Sy2s, with sources having a stronger reflection component being brighter and thus more easily detected. However, the difference between the average luminosities of MOB ($L^{MOB} \simeq 1.7 \times 10^{43} \rm \,erg \,s^{-1}$) and LOB Sy2s ($L^{LOB} \simeq 2.1 \times 10^{43} \rm \,erg \,s^{-1}$) is insignificant. Moreover, a KS test results in a probability of $\simeq 86\%$ that the two samples of Seyfert\,2s are statistically compatible.
Different inclination angles might also play a role in explaining the large reflection of MOB Sy2s, although the average values of $i$ would have to be extreme to account for the observed differences.

A possible explanation is that in MOB Sy2s the putative torus, which is also the dominant reflector, covers a larger fraction of the X-ray source than in lightly obscured objects. A similar geometry was hypothesized by Ueda et al. (2007) to explain the {\it Suzaku} spectra of SWIFT\,J0601.9$-$8636 and SWIFT\,J0138.6$-$4001. Further evidence for the existence of these deeply buried AGN was found by Eguchi et al. (2009).
Being more absorbed, MOB Sy2s might on average have more matter surrounding the active nucleus, which would be responsible for the larger reflection. 
Ramos Almeida et al. (2009, 2011) analyzing the mid-infrared emission of Seyfert galaxies found that the clumpy absorbers in Sy2s have larger covering factors ($C_{\rm\,T}=0.95\pm 0.02$) than those in Sy1s ($C_{\rm\,T}=0.5\pm 0.1$). Among the 12 Sy2s in their sample, ten have $N_{\rm\,H}\gtrsim 10^{23}\rm\,cm^{-2}$, which supports our argument of a larger average covering fraction of the absorber for more obscured sources.
The small value of the average reflection parameter of Sy1s and Sy1.5s might imply that on average the reflection from the disk does not play as important a role as that of the absorber.

A value of $R>1$ is unphysical if related only to the geometry, and might imply that part of the direct emission is blocked by partially covering material, with the transmission efficiency being $\lesssim 1/R$ (Ueda et al. 2007). If the X-ray source is partially covered by Compton-thick material, the reflected component would in fact have a stronger relative influence over the continuum. 
Krolik \& Begelman (1988) showed that a torus with a smooth dust distribution cannot survive close to the AGN, and proposed that the material is distributed in a clumpy structure. Mid-infrared spectra of Seyfert galaxies have been proven to be consistent with the clumpy torus scenario (e.g., Mor et al. 2009).
Evidence has been found in the past few years of significant variations in $N_{\rm \,H}$, with changes from Compton-thick to Compton-thin states on timescales from weeks (Risaliti et al. 2005) to $\sim$\,10 hours (Risaliti et al. 2009). These variations have been interpreted as eclipses of the X-ray source caused by the BLR. If the BLR lies between the X-ray source and the putative torus, then it could provide the CT clumps needed to deplete the continuum emission.
The variations observed in NGC\,1365 (Risaliti et al. 2005, 2009) might be common in MOB Sy2s, and the extremely long {\it INTEGRAL} IBIS/ISGRI observations we used to derive the average spectra might have registered some of them. 
The presence of CT clumps might be a common characteristic in the absorbers of Seyfert galaxies, and be responsible, at least in part, for the large reflection observed in MOB Sy2.

Both effects might be at work in MOB Sy2s, with more absorbed objects having on average more matter around the X-ray source, and the distribution of this matter being clumpy.

\subsection{Seyfert\,2s with and without PBLs}
Spectropolarimetric surveys indicate that only 30--50\% Seyfert\,2s show PBLs (Tran 2001, 2003). The reason for the non-detection of PBLs in all Seyfert\,2s is still debated, and it has been hypothesized (Tran 2001, 2003) that NPBL Sy2s might be a different class of Seyfert\,2, that possibly lack the BLR and have large-scale characteristics more similar to Seyfert\,1s. Deluit (2004) discussed the possible existence of differences in the 15--136 keV spectra of Seyfert\,2s with and without PBLs using {\it BeppoSAX} data, and concluded that the characteristics of PBL Sy2s appear to be more similar to those of Seyfert\,1s than NPBL Sy2s.
Using a slightly larger sample, but broader energy range and much longer exposures, we analyzed the stacked spectra of PBL and NPBL Sy2s, finding that the two samples do not show any significant spectral differences. The greatest difference between the two samples is their luminosities. The average 17--80 keV luminosity of PBL Sy2s ($L^{\rm\,PBL} \simeq 1.9 \times 10^{43} \rm \,erg \,s^{-1}$) is greater by a factor of $\simeq 3$ than that of NPBL Sy2s ($L^{\rm\,NPBL} \simeq 6 \times 10^{42} \rm \,erg \,s^{-1}$). 
The luminosity of the $[\rm\,OIII]$ line is often used as an indicator of the bolometric luminosity of AGN. Collecting data from the literature, we found that the average luminosity of PBL Sy2s ($L_{[\rm\,OIII]}=4.2\times10^{40} \rm \,erg\,s^{-1}$) is still higher than that of NPBL Sy2s ($L_{[\rm\,OIII]}=10^{40} \rm \,erg\,s^{-1}$). Although this does not give any indication of the intrinsic luminosity of the BLR, it implies that the bolometric luminosity of PBL Sy2s is on average greater than that of NPBL Sy2s, confirming what was found by Tran (2001) and Lumsden \& Alexander (2001). We also found that in our sample NPBL Sy2s are on average slightly more absorbed ($N_{\rm \,H}^{\rm\,NPBL}=3 \times 10^{23}\rm \,cm^{-2}$) than PBL Sy2s  ($N_{\rm \,H}^{\rm\,PBL}=1.5 \times 10^{23}\rm \,cm^{-2}$).

The lack of remarkable differences in the hard X-ray spectra of PBL and NPBL Seyfert\,2s fits the UM, according to which they are supposed to belong to the same family.

\subsection{Narrow-line Seyfert\,1s}
The 0.1--10 keV spectra of NLS1s are steeper than those of their broad-line counterparts (e.g., Netzer 2001). Owing to this, few of them are detected above 20 keV, and only a handful of NLS1s have been studied so far in the hard X-rays.  A sample of five objects was examined using IBIS/ISGRI data (up to 100 keV) by Malizia et al. (2008), who found that their spectra present a continuum significantly steeper than Seyfert\,1s and have an average photon index of $\Gamma =2.6\pm 0.3$. They interpreted this as being due to a low energy ($E_C\leq 60$\,keV) of the cutoff.
Our sample of 205 Seyfert galaxies contains 14 NLS1s, which makes it the largest ever studied at these energies. In our sample, NLS1s represent $\simeq 15\%$ of the total number of Sy1s and Sy1.5s. This is a much smaller fraction than the $\sim 50\%$ of {\it ROSAT} soft X-ray selected NLS1s (Grupe 2004), but is consistent with the $\sim15\%$ obtained from an optically selected sample (Williams et al. 2002).   
From our analysis, we confirmed that in the hard X-rays the average spectrum of NLS1s is also steeper than those of Sy1s and Sy1.5s, with the spectra clearly diverging above $\sim\,40$\,keV. This might be due to different values of $\Gamma$ or to different energies of the cutoff. 
The fact that up to $\sim\,40$\,keV the average spectrum of NLS1s is consistent with that of Sy1s favors the hypothesis that the photon indices are consistent, while the energy of the cutoff is lower for NLS1s. This is supported also by the fact that for the objects for which good quality data are available in the soft X-ray band, the values of the photon indices are not significantly different from those of Sy1s (Malizia et al. 2008, Jim\'{e}nez-Bail\'{o}n et al. 2008, Winter et al. 2009). Consistent photon indices might be due to a bias towards low values of $\Gamma$ for our hard X-ray selected sample.

\subsection{The unified model of AGN}

The fact that the average primary (i.e. excluding the reflected component) hard X-ray emission of Seyfert galaxies is independent of their optical classification, and that PBL and NPBL Sy2s have consistent hard X-ray spectra fits well the zeroth order UM. However, the idea that the anisotropic absorber is the same for all classes does not allow us to easily explain the large reflection seen in Seyfert\,2s, and in particular in MOB Seyfert\,2s. 

A possible geometry might be that in MOB Sy2s the absorbing material covers a large fraction of the X-ray source, which is seen through clumps of Compton-thick material, possibly located in the Compton-thin absorber ($N_{\rm \,H}\simeq 10^{23} \rm \,cm^{-2}$). This geometry would also explain the low hard X-ray to $[\rm\,OIV]$ ratio found in Compton-thin Seyfert\,2s by Rigby et al. (2009), and why the luminosity of the Compton-thin Sy2s of our sample is significantly lower than that of Sy1s and Sy1.5s.

\subsection{The cosmic X-ray background}
The CXB is known to be produced by the integrated emission of unresolved AGN. Thus, to fully understand the spectrum of the CXB, it is extremely important to study the average hard X-ray spectra of Seyfert galaxies. 
In their CXB synthesis model, Gilli et al. (2007) assumed the disk to be the main reflector, considering a weaker reflection for Seyfert\,2s than for Seyfert\,1s, and a value of $R=0.37$ for the unabsorbed reflector. Using these parameters, they hypothesized that, to explain the great amount of flux around $\sim30$\,keV, CT Sy2s contribute up to $\sim\,30\%$ of the CXB. Risaliti et al. (1999) and Guainazzi et al. (2005) estimated that $\sim50\%$ of all obscured AGN are CT. However, so far this large amount of CT AGN has not been observed (e.g., Ajello et al. 2008a, Paltani et al. 2008), and only few of them are known. Treister et al. (2009) (hereafter T09) discussed the strong degeneracy between the fraction of CT AGN and the value of the reflection parameter used, finding that for a value of $R\simeq 1$, a lower fraction of CT objects ($\sim10\%$) is needed. Gandhi et al. (2007) also showed how an enhanced reflection component, due to light bending in their case, might decrease the large fraction of heavily absorbed AGN.

We have shown here that the average spectrum of MOB Sy2s has a strong reflection component, and an average value of $R\gtrsim 1$, although for LOB objects we found a value of $R$ smaller than that used by Gilli et al. (2007). This might imply that obscured objects start to contribute significantly to the peak of the CXB already for $N_{\rm\,H}\sim10^{23}\rm\,cm^{-2}$. 
To test this, we calculated the average value of $R$ for AGN, and extrapolated the CT AGN density as shown in Fig.\,4 of T09.
To obtain the average value of $R$, we used the $N_{\rm\,H}$ distribution of T09, which considers about the same amount of absorbed (AB, $N_{\rm\,H} \geq 10^{23}\rm\,cm^{-2}$, which includes MOB and CT AGN) and LOB ($N_{\rm\,H} < 10^{23}\rm\,cm^{-2}$) objects. We showed that the average spectra of CT and MOB Sy2s are consistent, thus we divided the AGN into two classes, AB and LOB, and used their average reflection parameters (Sect.\,\ref{Sy_spanalysis}) to calculate the average value of $R$. The value obtained $\bar{R}\sim1.1$ agrees with that found by T09, and implies that CT AGN have a local density of $\sim 2\times 10^{-6} \rm\,Mpc^{-3}$, and represent a fraction of $\sim 10\%$ of the total population of AGN.

We fitted the {\it INTEGRAL} IBIS/ISGRI CXB spectrum of T\"urler et al. (2010) with the sum of the best-fit models obtained for AB and LOB objects\footnote{In XSPEC, this is translated as \texttt{pexrav+pexrav}.}, fixing all the parameters to the best-fit values found in Sect.\,\ref{Sy_spanalysis}, leaving only the normalization free to vary. 
Following the $N_{\rm\,H}$ distribution of T09, and assuming that all the objects contribute equally to the CXB, we forced the two values of the normalization to have the same value ($A_{LOB}=A_{AB}$). We set the redshift to $z=1$. We thus obtained a fit with only one free parameter, the global normalization. This simplified model yields a good fit of the CXB spectrum ($\chi^2=13.5$ for 15 DOF) in the 20--200\,keV band.

\begin{figure}[h]
\centering
\includegraphics[width=9cm]{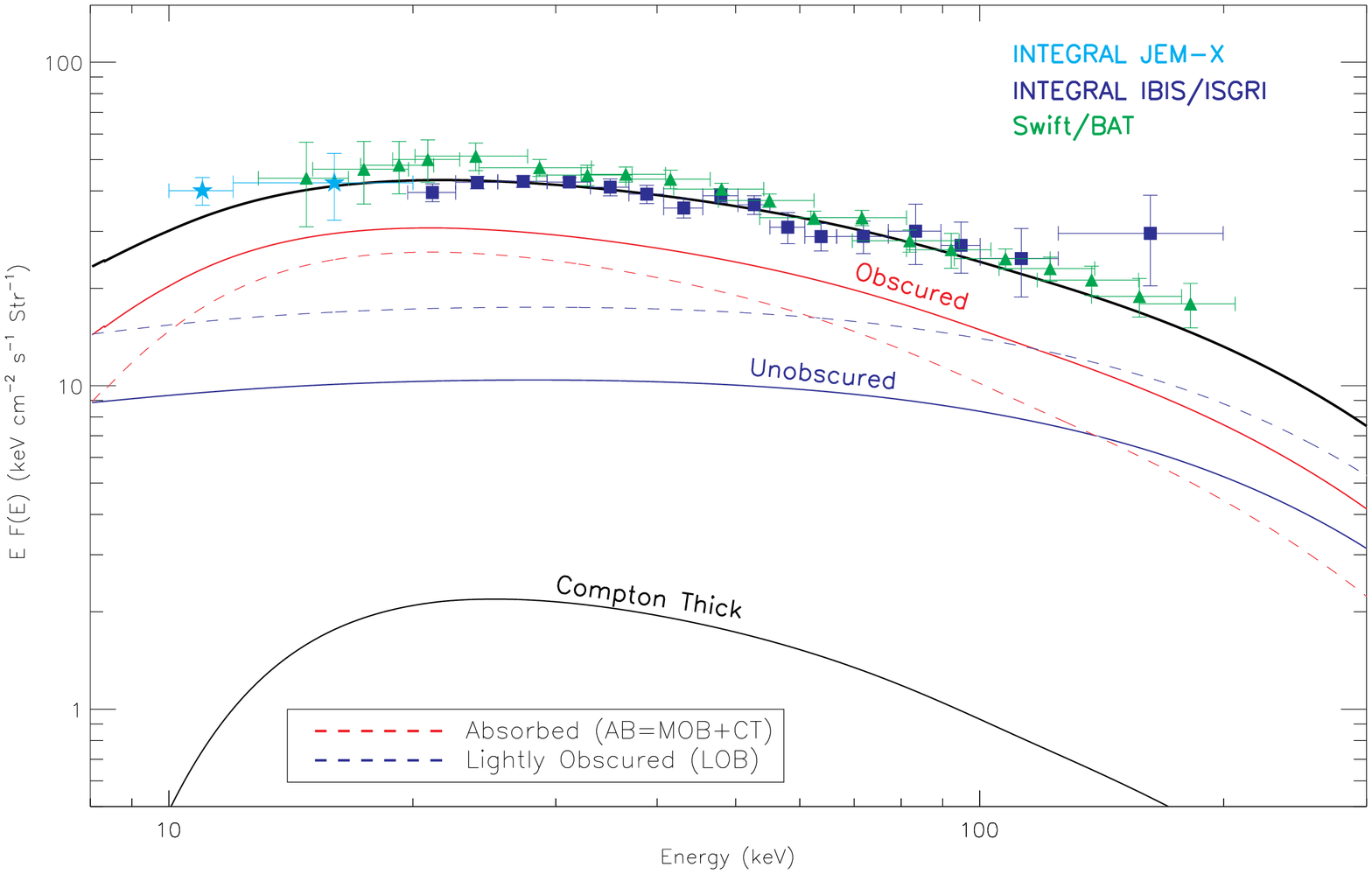}
\caption{Fit to the spectrum of the CXB measured by {\it INTEGRAL} IBIS/ISGRI (squares, T\"urler et al. 2010). The model used is a combination of our best-fit models for the average spectra of the different classes of objects (AB and LOB), normalizing their ratio using the $N_{\rm \,H}$ distribution of Treister et al. (2009). We also show the data obtained by {\it INTEGRAL} JEM-X (stars, T\"urler et al. 2010) and {\it Swift} BAT (triangles, Ajello et al. 2008b) above 10\,keV.}
\label{fig:cxb_model}
\end{figure}

In Fig.\ref{fig:cxb_model}, we show our fit to the CXB, and the contribution of the different components to the spectrum. Using the $N_{\rm\,H}$ distribution of T09, we transformed the models of the average spectra of LOB and AB objects into those of the three classes used by Gilli et al. (2007): obscured ($10^{24}\geq N_{\rm\,H}\geq 10^{22}\rm\,cm^{-2}$), unobscured ($N_{\rm\,H}< 10^{22}\rm\,cm^{-2}$), and CT objects. We added Compton scattering and photoelectric absorption to the model of CT AGN (setting $N_{\rm\,H}= 10^{24}\rm\,cm^{-2}$).

\section{Summary and conclusion}	\label{sect:concl}	
We have presented a study of the average hard X-ray spectra of Seyfert galaxies performed using {\it INTEGRAL} IBIS/ISGRI data. In the following, we summarize our findings:
\begin{itemize}
\renewcommand{\labelitemi}{$\bullet$}
\smallskip
\item  Different classes of Seyfert galaxies have the same average nuclear continuum ($\Gamma\simeq 1.8$, $E_C\gtrsim 200$\,keV), and their Comptonizing medium has on average a consistent value of the optical depth ($\tau\simeq 0.8$).
\smallskip
\item NLS1s have a steeper hard X-ray spectrum than Seyfert\,1s and Seyfert\,1.5s above $\sim 40$\,keV. This is likely due to a lower cutoff energy, as proposed by Malizia et al. (2008).
\smallskip
\item PBL and NPBL Sy2s have similar hard X-ray spectra, the only difference between the two classes in our incomplete sample being their luminosity.
\smallskip
\item Seyfert\,2s have on average a stronger reflection component than Seyfert\,1s and Seyfert\,1.5s. 
\smallskip
\item Most of the reflection of Seyfert\,2s comes from MOB ($10^{23} \rm \,cm^{-2} \leq N_{\rm \,H}< 10^{24} \rm \,cm^{-2}$) Sy2s.
\smallskip
\item The large amount of reflection observed in MOB Seyfert\,2s might be explained by an X-ray source highly covered by the absorbing material. The Compton-thin absorber might contain CT clumps, partially covering the X-ray source. 
\smallskip
\item The large amount of reflection of MOB Sy2s reduces by a factor of $\sim 3$ the amount of CT Seyfert\,2 needed to explain the CXB peak. Our results are consistent with the fraction of CT AGN being $\sim 10\%$.
\smallskip
\end{itemize}
Our results, while confirming that the X-ray engine is the same for all classes of Seyfert galaxies, point towards the existence of significant differences in the structure of the medium surrounding the X-ray source between LOB Seyfert galaxies and MOB Seyfert\,2s. These differences cannot be solely due to viewing angle dependence, but might imply that there are morphological differences between LOB and MOB AGN.


\appendix
\section{List of sources}\label{list_sources}
In the following, we list the coordinates, redshifts, detection significances, count rates, column densities, and luminosities of the samples of Seyfert\,1s (Table\,\ref{tab:sy1}), Seyfert\,1.5s (\ref{tab:syInt}), Compton-thin Seyfert\,2s (\ref{tab:sy2}), CT Seyfert\,2s (\ref{tab:CT}), and NLS1s (\ref{tab:NLS1}) used for the stacking analysis.

\begin{table*}[t]
\caption{Coordinates, redshifts, detection significances, count rates, column densities, and luminosities of the Seyfert\,1s used for the stacking analysis. Detection significances, count rates, and luminosities are in the 17--80 keV band. The flux of the Crab in the 17--80 keV band is of $278\rm \,ct\,s^{-1}$.}
\begin{center}
\label{tab:sy1}
\resizebox{0.9\textwidth}{!}{%
\begin{tabular}[c]{lccccccc}
\hline \hline \noalign{\smallskip}
Name&RA& Dec& $z$& Det. significance & Count rate & $N_{\rm \,H}$ & $\log L_{17-80\rm \,keV}$\\
              &[{\tiny h m s}]&[{\tiny $^\circ$ ' ''}]&   & [{\tiny $\sigma$}] & [{\tiny $\rm \,ct \,s^{-1}$}] & [{\tiny $10^{22} \rm \,cm^{-2}$}]& [\tiny {$\rm \,erg \,s^{-1}$}] \\
\noalign{\smallskip\hrule\smallskip}
IGR J02086$-$1742		&    	02 08 31.4    &      $-$17 43 04.8 &	0.129& 7.3 &  $ 0.31 \pm 0.04 $ & $ <0.017^{a} $ &        44.78 \\
IGR J02097+5222   &    02 09 37.7   &    $+$52 26 43.6    & 0.049	& 13.5 &  $ 0.56 \pm 0.04 $ & $  0.03^{b} $ &                 44.20 \\
Mrk 590 & 02 14 33.6 & $-$00 46 00.3 & 0.027  & 5.0 &  $ 0.13 \pm 0.03 $ & $  0.03^{c} $ &                                    43.05  \\  
Mrk 1040   &    02 28 14.6   &    +31 18 39.4            &	0.016 & 8.4 &  $ 0.72 \pm 0.09 $ & $ 0.067^{d} $ &                43.33 \\
RBS 345   &    02 42 16.0   &    +05 31 48.0                  &	0.069 &  5.1 &  $ 0.23 \pm $ 0.04& --  &                      44.11 \\
ESO 548$-$01		&    	03 41 54.7    &      $-$21 15 28.8	   & 0.014	& 6.1  &  $ 0.64 \pm 0.11$& -- &                      43.17 \\ 
UGC 3142   &    04 43 46.9   &    +28 58 19.0            &	0.022 & 13.8 &  $ 0.58 \pm 0.04$& $ 1.4^{e} $ &                   43.52 \\
LEDA 168563   &    04 52 04.7   &    +49 32 45.0            & 0.029	& 11.6 &  $ 0.51 \pm 0.04 $ & $  <0.22^{b} $ &            43.70 \\
Ark 120   &    05 16 11.5   &    $-$00 09 00.6             & 0.033	& 29.3 &  $ 0.77 \pm 0.03 $& $  <0.1^{c} $ &              43.99 \\
EXO 055620$-$3820.2   &    05 58 02.0   &    $-$38 20 01.0      & 0.034	& 5.6 &  $ 0.31 \pm 0.06 $ & $ 2.57^{f} $ &           43.62 \\
IRAS 05589+2828   &    06 02 09.7   &    +28 28 17.0        & 0.033	& 17.5 &  $ 0.61 \pm 0.03 $ & $ <0.04^{b} $ &             43.89 \\
SWIFT J0640.4$-$2554	&    	06 40 10.8    &      $-$25 49 48.0 & 0.026	& 5.0 &  $ 0.39 \pm 0.08 $ & $  0.33^{f} $ &          43.49                            \\		
IGR J07597$-$3842   &    07 59 41.7   &    $-$38 43 57.4    & 0.040	& 21.8 &  $ 0.60 \pm 0.03 $ & $  0.05^{d} $ &             44.05                         \\
PG 0804+761   &    08 10 58.7   &    +76 02 42.5         &	 0.100 & 5.0  &  $ 0.20 \pm 0.04 $& $ 0.023^{d} $ &               44.37                        \\
Fairall 1146 &    08 38 30.7 &    $-$35 59 35.0 & 0.032  & 14.0 &  $ 0.34 \pm 0.02 $ & $  0.1^{d} $ &                         43.61             \\   				
IGR J09026$-$4812   &    09 02 37.3   &    $-$48 13 34.1     & 0.039	& 21.6  &  $ 0.37 \pm 0.02 $& $  0.9^{g} $ &              43.82                         \\
SWIFT J0917.2$-$6221   &    09 16 09.4   &    $-$62 19 29.5  & 0.057	& 12.1 &  $ 0.34 \pm 0.03 $ & $  0.5^{h} $ &              44.11                        \\
4U 0937$-$12   &    09 45 42.1   &    $-$14 19 35.0          & 0.008	& 17.1 &  $ 0.99 \pm 0.06 $& $  1.19^{f} $ &              42.87                        \\
SWIFT J1038.8$-$4942 &    10 38 45.0 &    $-$49 46 55.0 & 0.060  & 9.3 &  $ 0.25 \pm 0.03 $ & $  1.55^{f} $ &                   44.02                   \\   
IGR J11457$-$1827   &    11 45 41.0   &    $-$18 27 29.0          & 0.033 	& 6.8 &  $ 0.87 \pm 0.13 $& $ \simeq 0^{f} $ &    44.05                                     \\
IGR J12107+3822    &    12 10 43.4   &    +38 22 51.6     & 0.023	& 5.3 &  $0.17 \pm 0.03$ & -- &                           43.02            \\
IGR J12136$-$0527   &    12 13 37.0   &    $-$05 26 53.0          & 0.066	& 5.3 &  $ 0.16 \pm 0.03 $ & $  2.0^{i} $ &           43.91                            \\
IGR J12172+0710   &    12 17 09.0   &    +07 09 33.0          & 0.007	& 8.3 &  $ 0.22 \pm 0.03 $ & $  0.15^{i} $ &          42.10                             \\
Mrk 50   &    12 23 24.1   &    +02 40 44.8              & 0.023	& 8.9  &  $ 0.22 \pm 0.02$ & $  0.018^{d} $ &             43.14                          \\
NGC 4593   &    12 39 39.4   &    $-$05 20 39.3            & 0.009	& 39.7 &  $ 1.03 \pm 0.03 $ & $  0.02^{j} $ &             42.99                          \\
ESO 323$-$77   &    13 06 26.6   &    $-$40 24 50.0             & 0.015	& 16.6 &  $ 0.50 \pm 0.03 $ & $ 6.0 ^{k} $ &          43.12                              \\
IGR J13109$-$5552   &    13 10 43.1   &    $-$55 52 11.7     & 0.104	& 15.2 &  $ 0.31 \pm 0.02 $ & $ <0.1 ^{c} $ &             44.60                           \\
Mrk 279 & 13 53 03.5 & +69 18 29.2 & 0.031  & 5.0 &  $ 0.64 \pm 0.13 $ & $ 0.02 ^{f} $ &                                      43.86       \\ 
RHS 39   &    14 19 22.2   &    $-$26 38 41.0                 & 0.022	& 10.4 &  $ 0.46 \pm 0.04 $& $ <0.05^{c} $ &          43.42                             \\
IGR J14471$-$6414   &    14 46 28.3   &    $-$64 16 24.1     & 0.053	& 9.7 &  $ 0.19 \pm 0.02 $ & $ <0.1^{c} $ &               43.80                       \\
IGR J16119$-$6036   &    16 11 51.4   &   $-$60 37 53.1     &  0.016	& 14.8 &  $ 0.33 \pm 0.02 $& $  0.1^{d} $ &               43.00                         \\
IGR J16482$-$3036   &    16 48 14.9   &    $-$30 35 06.1     & 0.031	& 29.9 &  $ 0.62 \pm 0.02 $& $  0.13^{c} $ &              43.85                          \\
IGR J16558$-$5203   &    16 56 05.7   &    $-$52 03 41.2    &  	0.054	&  23.9 &  $ 0.47 \pm 0.02 $& $ 0.011^{c} $ &         44.21                              \\
IGR J17418$-$1212   &    17 41 55.3   &    $-$12 11 57.5      & 0.037	& 19.4 &  $ 0.43 \pm 0.02 $& $  0.1^{d} $  &              43.84                         \\
IGR J18027$-$1455   &    18 02 48.0   &     $-$14 54 54.8            & 0.035	& 23.6 &  $ 0.42 \pm 0.02 $ & $  0.41^{l} $  &    43.78                                  \\
IGR J18259$-$0706   &    18 25 57.6   &   $-$07 10 22.8     & 0.037	& 13.2 &  $ 0.24 \pm 0.02 $ & $  0.6^{c} $ &              43.58                         \\
ESO 140-43   &    18 44 54.0&    $-$62 21 52.9                  & 0.014	& 11.8 &  $ 0.52 \pm 0.04 $ & $ 1.8^{e} $ &           43.08                           \\        
IGR J18559+1535   &    18 56 00.6   &    +15 37 58.0        & 0.084	& 18.3 &  $ 0.35 \pm 0.02 $ & $  0.7^{a} $  &             44.46                         \\
ESO 141-55   &    19 21 14.2   &    $-$58 40 15.0             & 0.037	& 12.7 &  $ 0.64 \pm 0.05 $ & $ 0.004^{c} $  &        44.01                              \\
SWIFT J1933.9+3258 & 19 33 47.3 & +32 54 25.0 & 0.056  & 7.0 &  $ 0.22 \pm 0.03 $ & $  <0.04^{m} $  &                         43.91             \\ 
IGR J19405$-$3016   &    19 40 15.2   &    $-$30 15 48.5     & 0.052	& 9.0 &  $ 0.29 \pm 0.03 $ & $ <0.1^{c} $  &              43.97                        \\
RX J2135.9+4728   &    21 35 54.4   &    +47 28 28.3     & 0.025	& 14.1 &  $ 0.28 \pm 0.02 $ & $  0.4^{c} $  &             43.31                         \\
IGR J22292+6647   &    22 29 13.5   &    +66 46 51.8     & 0.113	& 9.7 &  $ 0.22 \pm 0.02 $ & $  0.2^{c} $  &              44.52                        \\
IGR J23206+6431   &    23 20 36.8   &    +64 30 42.8      & 0.072	& 9.0 &  $ 0.15 \pm 0.02 $ & $  0.6^{c} $  &              43.96                        \\
\hline
\multicolumn{8}{l}{{\bf Notes.} {\footnotesize $^a$ Rodriguez et al. (2010); $^b$ Winter et al. (2008) and references therein; $^c$ Beckmann et al. (2009) and ref. therein;}} \\
\multicolumn{8}{l}{{\footnotesize  $^d$ Bodaghee et al. (2007) and ref. therein; $^e$ Ricci et al. (2010); $^f$ Winter et al. (2009); $^g$ Tomsick et al. (2008); $^h$ Malizia }}\\
\multicolumn{8}{l}{{\footnotesize  et al. (2007); $^i$ Paltani et al. (2008) and ref. therein; $^j$ Beckmann et al. (2006) and ref. therein; $^k$ Jim\'{e}nez-Bail\'{o}n }}\\
\multicolumn{8}{l}{{\footnotesize  et al. (2008); $^{l}$ Panessa et al. (2008);   $^{m}$ Landi et al. (2007a). }}\\
\end{tabular}
}
\end{center}
\end{table*}

\begin{table*}[t]
\caption{ Coordinates, redshifts, detection significances, count rates, column densities, and luminosities of the Seyfert\,1.5s used for the stacking analysis. Detection significances, count rates, and luminosities are in the 17--80 keV band.  The flux of the Crab in the 17--80 keV band is of $278\rm \,ct\,s^{-1}$.}
\begin{center}
\label{tab:syInt}
\resizebox{0.9\textwidth}{!}{%
\begin{tabular}[c]{lccccccc}
\hline \hline \noalign{\smallskip}
Name&RA& Dec& $z$ & Det. significance & Count rate & $N_{\rm \,H}$& $\log L_{17-80\rm \,keV}$\\
              &  [{\tiny h m s}]&[{\tiny $^\circ$ ' ''}]&   &  [{\tiny $\sigma$}] &  [{\tiny $\rm \,ct \,s^{-1}$}]  & [{\tiny $10^{22} \rm \,cm^{-2}$}] & [\tiny {$\rm \,erg \,s^{-1}$}] \\
\noalign{\smallskip\hrule\smallskip}
  				 		
IGR J00335+6126 &    00 33 18.4 &    +61 27 42.4 & 0.105  & 9.2  &  $ 0.14 \pm 0.02 $ & $  0.5^{a} $   &                     44.26            \\ 
Mrk 1152 &  01 13 50.1 & $-$14 50 44.1 & 0.052  & 5.0 &  $ 0.47 \pm 0.09 $& $  2.7^{b} $    &                                44.17 \\ 
NGC 985 &    02 34 37.8 &    $-$08 47 15.4 &   0.043   & 14.9  &  $ 0.42 \pm 0.03 $& $ 0.6 ^{c} $    &                       43.96          \\ 
IGR J03334+3718 &    03 33 18.8 &    +37 18 11.4 &  0.055   &  5.0 &  $ 0.18 \pm 0.03 $& --    &                             43.81    \\ 			
4U 0517+17 &    05 10 45.5 &    +16 29 55.0 & 0.017  & 25.3  &  $ 0.90 \pm 0.04 $& $  0.1^{c} $    &                         43.40        \\	
SWIFT J0519.5$-$3140 &    05 19 35.7 &    $-$32 39 25.0 & 0.013   & 14.9  &  $ 0.67 \pm 0.05 $& $ 26.6 ^{e} $    &             43.12                    \\   					
MCG+08$-$11$-$011 &    05 54 53.6 &    +46 26 21.8 &  0.020   & 24.9  &  $ 1.81 \pm 0.07 $& $  0.18^{c} $    &                   43.93              \\   			
Mrk 6 &    06 52 12.3 &    +74 25 36.8 &  0.019   &  16.7 &  $ 0.74 \pm 0.04 $& $  10^{d} $    &                             43.50    \\  		
ESO 209$-$12 &    08 01 57.6 &    $-$49 46 42.0 & 0.040   & 15.8  &  $ 0.34 \pm 0.02 $& $  0.1^{c} $    &                      43.81           \\  					 		
IGR J09253+6929 &    09 25 47.6 &    +69 27 51.5 & 0.039    & 5.1  &  $ 0.29 \pm 0.06 $& $  8^{f} $    &                     43.72            \\  			
NGC 3227 &    10 23 30.6 &    +19 51 53.7 &  0.003   &  17.4 &  $ 1.88 \pm 0.11 $& $ 6.8^{c} $    &                          42.30       \\  	
NGC 3516 & 11 06 47.6 & +72 34 08.3 & 0.009  & 6.5  &  $ 1.26 \pm 0.19 $& $  4^{g} $    &                                    43.08     \\ 
NGC 3783 &    11 39 01.8 &    $-$37 44 18.7 &  0.009  & 10.9  &  $ 2.05 \pm 0.19 $& $  0.08^{b} $    &                       43.29          \\  		
IGR J12415$-$5750 &    12 41 25.4 &    $-$57 50 04.0 & 0.024   &  16.2 &  $ 0.33 \pm 0.02 $ & $ <0.11^{c} $   &                43.35                 \\   			
IGR J13133$-$1109 &    13 13 17.0 &    $-$11 08 34.0 & 0.034   & 5.0  &  $ 0.20 \pm 0.04 $ &  --    &                          43.43       \\  		
MCG$-$06$-$30$-$015   &    13 35 53.8   &    $-$34 17 44.0          & 0.008	 &  23.2 &  $ 0.74 \pm 0.03 $ & $ 0.03 ^{c} $    &   42.75                              \\
IGR J14080$-$3023		&    	14 08 06.6  &      $-$30 23 52.6	&  0.023   & 5.0  &  $ 0.18 \pm 0.04 $ & $ 0.03 ^{h} $    &  43.05                               \\
NGC 5548 &    14 17 59.7 &    +25 08 13.4 &  0.017   &  7.1 &  $ 0.49 \pm 0.07 $ & $  0.51^{d} $    &                        43.22         \\  				
QSO B1419+480 &    14 21 29.6 &    +47 47 26.0 &   	0.072	   &  5.2 &  $ 0.25 \pm 0.05$ & $  0.5^{i} $    &                44.18                 \\				
PG 1501+106 &    15 04 01.2 &    +10 26 16.2 &  0.036    & 7.2  &  $ 0.49 \pm 0.07$ & $  3.9^{j} $    &                      43.87           \\ 						
SWIFT J1930.5+3414 &    19 30 13.7 &    +34 10 52.0 &  0.062   & 8.8  &  $  0.28 \pm 0.03$ & $ 27.5^{e} $    &               44.10                  \\   		
NGC 6814 &    19 42 40.4 &    $-$10 19 24.0 &   0.005    &  25.9 &  $ 0.79 \pm 0.03 $ & $ <0.05 ^{d} $    &                  42.37               \\
SWIFT J2009.0$-$6103 & 20 08 47.0 & $-$61 05 48.0 & 0.015   & 7.6 &  $ 0.61 \pm 0.08 $ & $ 4.53 ^{e} $     &                   43.21              \\ 
Mrk 509 &    20 44 09.8 &    $-$10 43 24.7 & 0.035   &  16.0 &  $ 1.12 \pm 0.07 $ & $ <0.01 ^{d} $     &                     44.21            \\   					
1RXS J211928.4+333259	&    	21 19 35.5  &       +33 33 03.6 & 0.051   & 9.8  &  $0.29 \pm 0.03 $ & --    &               43.95                  \\
1RXS J213944.3+595016	&    	21 39 42.7  &       +59 49 37.2 & 0.114  & 5.5  &  $ 0.13 \pm 0.02 $ & --    &               44.30                  \\
1H 2251$-$179   &    22 54 05.9   &   $-$17 34 55.3         & 0.064	 &  9.5 &  $ 1.12 \pm 0.12 $  & $ <0.19 ^{d} $    &      44.73                           \\
NGC 7469 &    23 03 15.8 &    +08 52 25.9 &  0.017  & 10.3  &  $ 0.78 \pm 0.08 $ & $ 0.061 ^{c} $    &                       43.42          \\ 
MCG$-$02$-$58$-$022 &    23 04 43.5 &   $-$08 41 08.6 & 0.047   & 8.2  &  $ 0.71 \pm 0.09 $ & $ <0.08 ^{d} $    &                  44.27               \\  	 		
	
\hline
\multicolumn{8}{l}{{\bf Notes.} {\footnotesize $^a$ Beckmann et al. (2009) and ref. therein; ; $^b$ Fiore et al. (1997); $^c$ Bodaghee et al. (2007) and ref. therein;  }}\\
\multicolumn{8}{l}{ {\footnotesize $^d$ Beckmann et al. (2006) and ref. therein; $^e$ Winter et al. (2009) and ref. therein; $^f$ Rodriguez et al. (2008); $^g$ Guainazzi}}\\
\multicolumn{8}{l}{ {\footnotesize  et al. (2002); $^h$ Landi et al. (2010); $^i$ Barcons et al. (2003); $^j$ Longinotti et al. (2010).}}\\

\end{tabular}
}
\end{center}
\end{table*}

\begin{table*}[t]
\caption{ Coordinates, redshifts, detection significances, count rates, column densities, and luminosities of the Seyfert\,2s used for the stacking analysis. Detection significances, count rates, and luminosities are in the 17--80 keV band.  The flux of the Crab in the 17--80 keV band is of $278\rm \,ct\,s^{-1}$.}
\begin{center}
\label{tab:sy2}
\resizebox{0.95\textwidth}{!}{%
\begin{tabular}[c]{lcccccccc}
\hline \hline \noalign{\smallskip}
Name&RA& Dec& $z$ & Det. significance & Count rate & $N_{\rm \,H}$& PBL & $\log L_{17-80\rm \,keV}$\\
              &  [{\tiny h m s}]&[{\tiny $^\circ$ ' ''}]&   &  [{\tiny $\sigma$}] &  [{\tiny $\rm \,ct \,s^{-1}$}]  & [{\tiny $10^{22} \rm \,cm^{-2}$}]&  & [\tiny {$\rm \,erg \,s^{-1}$}] \\
\noalign{\smallskip\hrule\smallskip}
IGR J00040+7020 &     00 04 01.5 &       +70 19 16.5 &   0.096       & 9.4  &  $ 0.18 \pm 0.02 $ & $3^{a} $   & -- &               44.29   \\
IGR J00254+6822 &     00 25 31.4 &       +68 21 41.3 &    0.012      & 12.2  &  $ 0.21 \pm 0.02 $ & $ 40^{b} $   & -- &            42.55  \\
Mrk 348  &     00 48 47.1 &       +31 57 25.1 &        0.014          &  23.2 &  $ 1.59 \pm 0.07 $ & $ 30^{b} $   &  Yes$^1$ &     43.56         \\
NGC 526	 &     01 23 54.2 &       $-$35 03 55.0 &       0.019          & 6.5  &  $ 0.69 \pm 0.11 $& $ 1.6^{c} $    &   -- &        43.47     \\
ESO 297$-$18 &     01 38 37.2 &       $-$40 00 40.7 &      0.025         & 9.6  &  $ 0.76 \pm 0.08 $ & $ 42^{b} $    &   -- &        43.75     \\
IGR J01528$-$0326 &     01 52 48.9 &       $-$03 26 51.0 &   0.017        &  9.1 &  $ 0.26 \pm 0.03$ & $ 14^{b} $   & -- &           42.95  \\
NGC 788	 &     02 01 06.5 &       $-$06 48 55.9 &       0.013           & 37.4  &  $ 0.94 \pm 0.03$ & $ <0.02^{b} $   & Yes$^2$ &  43.27           \\
Mrk 1018			&    	02 06 08.2  & $-$00 17 02.4 &  0.043   &  8.2 &  $ 0.22 \pm 0.03 $ & $ 0.03^{d} $    &   -- &          43.68   \\   				
IGR J02343+3229 &     02 34 19.9 &       +32 30 20.0 &     0.016        &  10.5 &  $ 0.75 \pm 0.07 $ & $ 2.2^{b} $   & -- &        43.35     \\
IGR J02501+5440	 &     02 50 41.8 &       +54 42 15.1 &  0.015         & 7.1  &  $ 0.29 \pm 0.04 $ & $< 70^{e}$   & -- &           42.89  \\
IGR J02524$-$0829	 &     02 52 23.3 &       $-$08 30 38.0 &     0.017        & 5.1  &  $ 0.18 \pm 0.04 $ & $ 12^{f} $    &   -- &    42.79         \\
NGC 1142 &     02 55 12.3 &       $-$00 11 01.7 &    0.029        & 25.0  &  $ 1.03 \pm  0.04$ & $ 45^{b} $    &  No$^3$ &         44.01    \\
NGC 1194 &     	03 03 46.1  &        $-$01 08 49.2 &   0.013   & 6.7  &  $ 0.34 \pm 0.05 $ & 1$^{g}$   & -- &                      42.83    \\
SWIFT J0318.7+6828 &     	03 19 02.4  &          +68 25 48.0  &  0.090   & 5.1  &  $ 0.24 \pm 0.05 $ & $ 4.1^{h} $   & -- &      44.36       \\
LEDA 15023 &     	04 23 48.2  &         +04 08 42.0 &    	0.046   & 8.4  &  $ 0.36 \pm 0.04 $ & --   & No$^1$ &                  43.95       \\
SWIFT J0444.1+2813 & 04 44 09.0 & +28 13 00.0 &  0.011 & 7.4 &  $ 0.33 \pm 0.04 $  & $ 3.39^{h} $   & -- &                         42.67       \\ 
SWIFT J0453.4+0404 &   04 53 31.0  &          +04 02 02.4 &  0.029    &  9.6 &  $ 0.27 \pm 0.03 $ & --    &   -- &                 43.43     \\
ESO 33$-$2	 &     04 55 59.6 &       $-$75 32 26.0 &       0.019          & 8.7  &  $ 0.35 \pm 0.04 $ & $ 0.1^{c} $    &   -- &   43.17          \\
SWIFT J0505.7$-$2348 &     05 05 45.7 &       $-$23 51 14.0 &   0.035       &  14.8 &  $ 0.86 \pm 0.06 $ & $ 6.57^{h} $    &   -- &  44.09           \\
NGC 2110 &     05 52 11.4 &       $-$07 27 22.4 &   0.007              & 48.5  &  $ 2.4 \pm 0.05 $ & $ 4.3^{b} $   & Yes$^2$ &     43.14        \\
IGR J06239$-$6052 &     06 23 45.6 &      $-$60 58 45.0 &    0.041         & 6.0  &  $ 0.31 \pm 0.05 $ & $ 20^{b} $   & -- &         43.78    \\
LEDA 96373 &     07 26 26.3 &       $-$35 54 21.0 &    0.029              &  5.9 &  $ 0.39 \pm 0.07 $ & $7^{i}$   & -- &           43.59      \\
IGR J07565$-$4139 &     07 56 19.6 &       $-$41 37 42.1 &   0.021       &  9.3 &  $ 0.23 \pm 0.02 $ & $ 1.1^{c} $   & -- &          43.08     \\
SWIFT J0920.8$-$0805 &     09 20 46.3 &       $-$08 03 21.9 &  0.020     & 6.8  &  $ 0.56 \pm 0.08 $ & $ 11.4^{h} $   & -- &         43.42     \\
MCG$-$05$-$23$-$016 &     09 47 40.2 &            $-$30 56 54.0 &       0.008  & 39.4 &  $ 2.16 \pm 0.06 $ & $ 1.6^{c} $ & Yes$^2$ &     43.21        \\
NGC 3081 &     09 59 29.5 &            $-$22 49 34.6 &   0.007            &  15.7 &  $ 0.95 \pm 0.06 $ & $ 66^{b} $   & Yes$^2$ &  42.74           \\
SWIFT J1009.3$-$4250 &     10 09 48.3 &            $-$42 48 40.4 & 0.032         &  15.6 &  $ 0.55 \pm 0.03 $ & $ 30^{b} $   & -- &  43.82           \\
IGR J10404$-$4625 &     10 40 22.3 &            $-$46 25 24.7 &  0.024       &  12.4 &  $0.45  \pm 0.04 $ & $ 3^{b} $   & -- &       43.48      \\
SWIFT J1049.4+2258 & 10 49 30.9 & +22 57 51.9 &  0.033 & 5.1 &  $ 0.42 \pm 0.08 $ & $ 42^{j} $   & -- &                            43.73    \\ 
SWIFT J1200.8+0650 &     12 00 57.9 &            +06 48 23.1 & 0.036     & 9.8  &  $ 0.32 \pm 0.03 $ & $9.3 ^{j} $    &   -- &     43.69        \\
IGR J12026$-$5349 &     12 02 47.6 &            $-$53 50 07.7 & 0.028         &  25.1 &  $ 0.57 \pm 0.02 $ & $ 2.2^{c} $   & -- &    43.72         \\
NGC 4074 &     	12 04 32.6  &       +20 14 56.4 & 0.022     & 5.9 &  $ 0.29 \pm 0.05 $  & $ 30^{h} $   &  No$^4$ &                 43.22   \\
NGC 4138 &     12 09 29.9 &            +43 41 06.0 &   0.003              & 8.6  &  $ 0.30 \pm 0.04 $ & $ 8^{b} $   & -- &         41.50    \\
NGC 4258 &     12 18 57.5 &            +47 18 14.3 &   0.002            & 5.5  &  $ 0.23 \pm 0.04 $ & $ 8.7^{b} $   & No$^5$ &     41.03        \\
NGC 4395 & 12 25 48.8 & +33 32 49     &  0.001   & 9.5  &  $ 0.29 \pm 0.03  $ & $ 0.15^{b} $   & -- & 40.53  \\
NGC 4507 &     12 35 36.5 &            $-$39 54 33.3 &   0.012              &  49.6 &  $ 2.29 \pm 0.05 $ & $ 29^{b} $   & Yes$^1$ & 43.59             \\
SWIFT J1238.9$-$2720 &     12 38 54.5 &            $-$27 18 28.0 &   0.024       & 11.5  &  $ 1.06 \pm 0.09 $ & $ 60^{b} $   & -- &  43.86           \\
IGR J12391$-$1612	 &     12 39 06.3 &            $-$16 10 47.1 &   0.036     &  11.2 &  $ 0.55 \pm 0.05 $ & $ 3^{b} $   & -- &       43.92      \\
IGR J12482$-$5828	&     	12 47 57.8  &             $-$58 29 59.1 & 0.028     & 6.7  &  $ 0.13 \pm 0.02 $ & $ 0.3^{i} $   & -- &     43.08        \\
ESO 323$-$32 &     12 53 20.4 &            $-$41 38 13.8 &  0.016         & 6.8  &  $ 0.23 \pm 0.03 $ & $ 7^{b} $    &   -- &        42.84     \\
NGC 4941 &     13 04 13.1 &            $-$05 33 05.7 &  0.003         &  6.5 &  $ 0.19 \pm 0.03 $ & $ 45^{k} $    &   No$^1$ &     41.30        \\
MCG$-$03$-$34$-$064 & 13 22 24.5 & $-$16 43 42.9 &  0.017 & 6.0 &  $ 0.40 \pm 0.07 $ & $ 41^{h} $    &   Yes$^2$ &                       43.13        \\ 
NGC 5252 &     13 38 16.0 &            +04 32 32.5 &    0.022  &  42.1 &  $ 1.40 \pm 0.03 $  & $ 0.68^{b} $   & Yes$^1$ &          43.90       \\
Mrk 268	 &     13 41 11.1 &            +30 22 41.2 &  0.041  &  8.4 &  $ 0.34 \pm 0.04 $ & $ 30^{e} $   & No$^4$ &                 43.83       \\
1AXG J135417$-$3746 &     	13 54 15.4  &    $-$37 46 33.0 &  0.052    & 8.3  &  $ 0.22 \pm 0.03 $ & $ 1.6^{l} $ &   -- &          43.85          \\
IGR J14175$-$4641	 &     14 17 03.9 &            $-$46 41 39.1 &  0.076        &  10.4 &  $ 0.24 \pm 0.02 $ & --   & -- &            44.21            \\ 
IGR J14471$-$6319 &     14 47 14.7 &            $-$63 17 19.6 &    0.038      &  8.5 &  $ 0.16 \pm 0.02 $ & $ 2^{b} $   & -- &       43.43                 \\
IGR J14561$-$3738 &     14 56 08.2 &           $-$37 38 53.8 &   0.024       &  7.9 &  $ 0.18 \pm 0.02 $ & --    &   -- &            43.09          \\
IGR J14579$-$4308 &     14 57 41.3 &            $-$43 07 57.0 &    0.016        &  17.1 &  $ 0.36 \pm 0.02 $ & $ 20^{b} $    &   -- & 43.04             \\
ESO 328$-$36 & 15 14 47.2 & $-$40 21 35 & 0.024 & 7.9  &  $ 0.17 \pm 0.02 	$ &  --   & -- &  43.06     \\
IGR J15161$-$3827 &     15 15 59.3 &            $-$38 25 48.3 & 0.036       & 5.0  &  $ 0.11 \pm 0.02 $ & $ 22^{m} $   & -- &        43.22          \\ 
NGC 5995 &     15 48 25.0 &            $-$13 45 28.0 &  0.025     &  8.3 &  $ 0.39 \pm 0.05 $ & $ 0.7^{b} $   & Yes$^1$ &          43.46          \\
IGR J15539$-$6142 &     15 53 35.2 &            $-$61 40 55.4 &  0.015          &  9.8 &  $ 0.21 \pm 0.02 $ & $ 18^{b} $    &   -- & 42.74            \\
SWIFT J1650.5+0434 &     	16 50 35.3  &         +04 35 42.0	&   0.032   &  5.4 &  $ 0.27 \pm 0.05 $ & --   & -- &              43.51            \\  
NGC 6300	 &     17 16 59.2 &            $-$62 49 11.0 &    0.003   & 32.4  &  $ 1.11 \pm 0.03 $ & $ 22^{b} $   & No$^1$ &       42.07      \\
IGR J17513$-$2011	 &     17 51 13.0 &           $-$20 12 14.5 &  0.047     &  22.3 &  $ 0.31 \pm 0.01 $ & $ 0.6^{b} $   & -- &       43.91      \\
IGR J18244$-$5622	 &     18 24 19.5 &            $-$56 22 08.7 & 0.017  & 14.4  &  $ 0.56 \pm 0.04 $ & $ 14^{b} $   & -- &           43.28         \\
Fairall 49 & 18 36 58.1 & $-$59 24 08.0 & 0.019  & 5.2 &  $ 0.22 \pm 0.04  $ & $ 0.01^{n} $   & Yes$^2$ &                          42.97            \\ 
ESO 103$-$35	 &     18 38 20.3 &            $-$65 25 41.0 &   0.013    & 26.2  &  $ 1.19 \pm 0.05 $ & $ 21.6^{h} $   & -- &         43.37          \\
IGR J19077$-$3925	&     	19 07 38.6  &     $-$39 25 37.2 &  0.073     & 9.8  &  $ 0.27 \pm 0.03 $ & --   & -- &                     44.23          \\
IGR J19473+4452 &     19 47 19.4 &            +44 49 42.4 &  0.053       &  11.2 &  $ 0.33 \pm 0.03$ & $ 11^{b} $   & -- &         44.04        \\  
IGR J20216+4359	 &     20 21 48.1 &            +44 00 32.0 &   0.017          &  8.3 &  $0.17 \pm 0.02 $ & $ 13^{o} $   & -- &     42.76        \\
IGR J20286+2544	 &     20 28 34.9 &            +25 43 59.7 &  0.013       & 19.1  &  $ 0.66 \pm 0.03 $ & $ 42^{b} $    &   -- &    43.12         \\
Mrk 520		&     	22 00 30.2  &               +10 36 10.8 &  0.028    & 7.6  &  $ 0.75 \pm 0.10 $ & $ 2.4^{h} $    &   -- &      43.84       \\
NGC 7172 	&     22 02 01.7 &            $-$31 52 18.0 &  0.008      &  19.1 &  $ 1.31 \pm 0.07 $ & $ 9^{b} $    &  No$^1$ &      42.99       \\
NGC 7314 	&     22 35 46.1 &            $-$26 03 01.7 &  0.005     & 5.1  &  $ 0.43 \pm 0.09$ & $ 0.12^{b} $    &   Yes$^1$ &    42.10         \\
IGR J23308+7120 &     23 30 37.3 &            +71 22 44.8 &  0.037    & 5.0  &  $ 0.11 \pm 0.02 $ & $ 6^{b} $   & -- &             43.25        \\    
IGR J23524+5842	&     23 52 22.0 &            +58 45 31.5 &  0.162    &   11.5 &  $ 0.18 \pm 0.02 $ & $ 6^{b} $   & -- &           44.75        \\
\hline
\multicolumn{8}{l}{{\bf Notes.} {\footnotesize \textit{References for the column density}. $^a$ Landi et al. (2007b); $^b$ Beckmann et al. (2009) and ref. therein; $^c$ Bodaghee et al. (2007)}}\\
\multicolumn{8}{l}{ {\footnotesize  and ref. therein; $^d$ Pfeerkorn et al. (2001); $^e$ Masetti et al. (2008); $^f$ Rodriguez et al. (2010); $^g$ Greenhill et al. (2008); $^h$ Winter et al. (2009) }  }\\
\multicolumn{8}{l}{{\footnotesize and ref. therein;  $^i$ Landi et al. (2010); $^j$ Winter et al. (2008) and ref. therein; $^i$ Paltani et al. (2008); $^l$ Risaliti et al. (2000); } }\\
\multicolumn{8}{l}{{\footnotesize $^m$ Rodriguez et al. (2008); $^n$ Dadina et al. (2007); $^o$ Bikmaev et al. (2008).} }\\
\multicolumn{8}{l}{{\footnotesize \textit{References for the PBL}. $^1$ Shu et al. 2007 and ref. therein; $^2$ Veron-Cetty \& Veron 2010 and ref. therein; $^3$ Cai et al. 2010;} }\\
\multicolumn{8}{l}{{\footnotesize  $^4$ Nagao et al. 2000; $^5$ Barth et al. 1999.} }\\

\end{tabular}
}
\end{center}
\end{table*}

\begin{table*}[t]
\caption{ Coordinates, redshifts, detection significances, count rates, column densities, and luminosities of the Compton-thick Seyfert\,2s used for the stacking analysis. Detection significances, count rates, and luminosities are in the 17--80 keV band. The flux of the Crab in the 17--80 keV band is of $278\rm \,ct\,s^{-1}$.}
\begin{center}
\label{tab:CT}
\resizebox{0.9\textwidth}{!}{%
\begin{tabular}[c]{lccccccc}
\hline \hline \noalign{\smallskip}
Name&RA& Dec& $z$ & Det. significance & Count rate & $N_{\rm \,H}$& $\log L_{17-80\rm \,keV}$\\
              &  [{\tiny h m s}]&[{\tiny $^\circ$ ' ''}]&   &  [{\tiny $\sigma$}] &  [{\tiny $\rm \,ct \,s^{-1}$}]  & [{\tiny $10^{22} \rm \,cm^{-2}$}] & [\tiny {$\rm \,erg \,s^{-1}$}] \\
\noalign{\smallskip\hrule\smallskip}
NGC 1068 & 02 42 40.8 & $-$00 00 48.4 & 0.003  & 14.0 &  $ 0.45 \pm 0.03 $ & 150$^a$ &             41.91        \\ 
SWIFT J0601.9$-$8636 & 06 05 39.1 & $-$86 37 52.0 &  0.006 & 7.3 &  $ 0.40 \pm 0.06 $ & 101$^b$ &    42.40                 \\ 
Mrk 3 & 06 15 36.3 & +71 02 14.9 &  0.014 & 31.3 &  $ 1.43 \pm 0.05  $ & $110^c$ &                 43.70    \\ 
NGC 3281 & 10 31 52.1 & $-$34 51 13.3 & 0.011  & 12.0 &  $ 0.78 \pm 0.07 $ & $151^d$ &             43.26        \\ 
IGR J13091+1137 & 13 09 05.6 & +11 38 02.9 & 0.025  & 14.8 &  $ 0.51 \pm 0.03 $ & $90^c$ &         43.73            \\ 
NGC 5643 & 14 32 41.3 & $-$44 10 24.0 & 0.003  & 9.2 &  $ 0.20 \pm 0.02 $ & $60-100^e$ &           41.50          \\ 
Mrk 477 & 14 40 38.1 & +53 30 15.9 & 0.037  & 5.8 &  $ 0.44 \pm 0.07 $ & $>100^f$ &                44.09     \\ 
NGC 5728	& 	14 42 23.3  &   $-$17 16 51.6	& 0.009  & 13.6 &  $ 0.80 \pm 0.06 $ & $210^a$&		43.12		\\ 
IGR J16351$-$5806 & 16 35 13.2 & $-$58 04 49.7 &  0.009 & 12.9 &  $ 0.28 \pm 0.02 $ & $>150^g$ &     42.69                \\ 
NGC 7582 & 23 18 23.5 & $-$42 22 14.1 & 0.005  & 5.5 &  $ 1.20 \pm 0.20 $ & 160$^a$ &              42.77       \\ 
\hline
\multicolumn{8}{l}{{\bf Notes}. {\footnotesize $^a$ Della Ceca et al. (2008) and ref. therein; $^b$ Ueda et al. (2007);  $^c$Beckmann et al. (2006)  and ref.  therein;}}\\
\multicolumn{8}{l}{{\footnotesize  $^d$ Bodaghee et al. (2007) and ref. therein; $^e$ Guainazzi et al. (2004); $^f$ Bassani et al. (1999); $^g$ Malizia et al. (2009).}}\\
\end{tabular}
}
\end{center}
\end{table*}

\begin{table*}[t]
\caption{ Coordinates, redshifts, detection significances, count rates, column densities, and luminosities of the Narrow Lines Seyfert\,1s used for the stacking analysis. Detection significances, count rates, and luminosities are in the 17--80 keV band. The flux of the Crab in the 17--80 keV band is of $278\rm \,ct\,s^{-1}$.}
\begin{center}
\label{tab:NLS1}
\resizebox{0.9\textwidth}{!}{%
\begin{tabular}[c]{lccccccc}
\hline \hline \noalign{\smallskip}
Name&RA& Dec& $z$ & Det. significance & Count rate & $N_{\rm \,H}$& $\log L_{17-80\rm \,keV}$\\
              &  [{\tiny h m s}]&[{\tiny $^\circ$ ' ''}]&   &  [{\tiny $\sigma$}] &  [{\tiny $\rm \,ct \,s^{-1}$}]  & [{\tiny $10^{22} \rm \,cm^{-2}$}] & [\tiny {$\rm \,erg \,s^{-1}$}] \\
\noalign{\smallskip\hrule\smallskip}
1H 0323+342 & 03 24 41.2 & +34 10 45.9 & 0.063  &  5.9 &  $ 0.20 \pm 0.03 $ & 0.1$^a$ &                         43.97     \\
Mrk 110 & 09 25 12.9 & +52 17 10.5 & 0.035  &  5.0 &  $ 0.86 \pm 0.17 $ & $0.019^a$ &                           44.09   \\
NGC 4051 & 12 03 09.6 & +44 31 53.2 & 0.002  & 13.6  &  $ 0.47 \pm 0.03$ & $<0.02^b$  &                         41.34     \\
Mrk 766 & 12 18 26.6 & +29 48 45.6 & 0.013  &  7.7 &  $ 0.26 \pm 0.03 $ & $0.8^c$  &                            42.71  \\
NGC 4748			& 	12 52 12.0  &   $-$13 25 48.0	& 0.014  & 5.2  &  $ 0.16 \pm 0.03 $ & $\simeq 0^d$  &  42.57                            \\
Mrk 783 & 13 02 58.8 & +16 24 27.5 & 0.067  & 7.3  &  $ 0.25 \pm 0.03 $ & $0.046^a$ &                           44.12   \\
NGC 5506	 &     14 13 14.9 &            $-$03 12 27.0 & 0.007    &  47.2 &  $ 2.7 \pm 0.06 $ & $ 3.4^{b} $     &     43.19        \\
IGR J14552$-$5133 & 14 55 17.9 & $-$51 34 13.3 & 0.016  & 12.5  &  $ 0.25 \pm 0.02 $ & $0.1^a$  &                 42.88             \\ 
IRAS 15091$-$2107		& 	15 11 54.0  &   $-$21 21 21.6	&  	0.044   & 5.3  &  $ 0.22 \pm 0.04 $ &  $0.7^e$  &   43.70                           \\
IGR J16185$-$5928 & 16 18 26.4 & $-$59 28 45.3 & 0.035  & 10.7  &  $ 0.23 \pm 0.02 $ & $<0.1^f$  &                43.52              \\
IGR J16385$-$2057 & 16 38 31.1 & $-$20 55 25.0 & 0.027  & 7.7  &  $ 0.26 \pm 0.03 $ & $0.21^g$  &                 43.35             \\
IGR J19378$-$0617 &    19 37 33.1 &   $-$06 13 04.0 &    0.010   &  11.2 &  $ 0.32 \pm 0.03 $ & $ \simeq 0 ^{h} $    &         42.58                        \\			
ESO 399$-$20 & 20 06 57.2 & $-$34 32 54.0 & 0.025   &  6.5 &  $ 0.23 \pm 0.04 $ & $0.048^a$  &                    43.23          \\
IGR J21277+5656   &    21 27 45.9   &    +56 56 34.4      & 0.014	& 24.2 &  $ 0.52 \pm 0.02 $ & $  0.1^{a} $ &              43.08                         \\
\hline
\multicolumn{7}{l}{{\bf Notes.}{\footnotesize $^a$ Bodaghee et al. (2007) and ref. therein; $^b$ Beckmann et al. (2009) and ref. therein; } }\\
\multicolumn{7}{l}{{\footnotesize $^c$ Beckmann et al. (2006) and ref. therein;  $^d$ Grupe et al. (1998); $^e$ Jim\'{e}nez-Bail\'{o}n et al. (2008); }}\\
\multicolumn{7}{l}{{\footnotesize $^f$ Malizia et al. (2008); $^g$ Rodriguez et al. (2008),; $^h$ Malizia et al. (2008).}}\\
\end{tabular}
}
\end{center}
\end{table*}

\section{Spectra}\label{Sec:spectraIm}
\begin{figure*}[h!]
\centering
\begin{minipage}[!b]{.48\textwidth}
\centering
\includegraphics[height=9cm,angle=270]{16409_sy1s.eps}
\end{minipage}
\hspace{0.05cm}
\begin{minipage}[!b]{.48\textwidth}
\centering
\includegraphics[height=9cm,angle=270]{16409_SyIs.eps} \end{minipage}
\begin{minipage}[!b]{.48\textwidth}
\centering
\includegraphics[height=9cm,angle=270]{16409_Sy2s.eps}\end{minipage}
\hspace{0.05cm}
\begin{minipage}[!b]{.48\textwidth}
\centering
\includegraphics[height=9cm,angle=270]{16409_CTsp.eps}\end{minipage}

\begin{minipage}[!b]{.48\textwidth}
\centering
\includegraphics[height=9cm,angle=270]{16409_NLsp.eps} \end{minipage}
\hspace{0.05cm}
\begin{minipage}[!b]{.48\textwidth}
\centering
\includegraphics[width=\columnwidth]{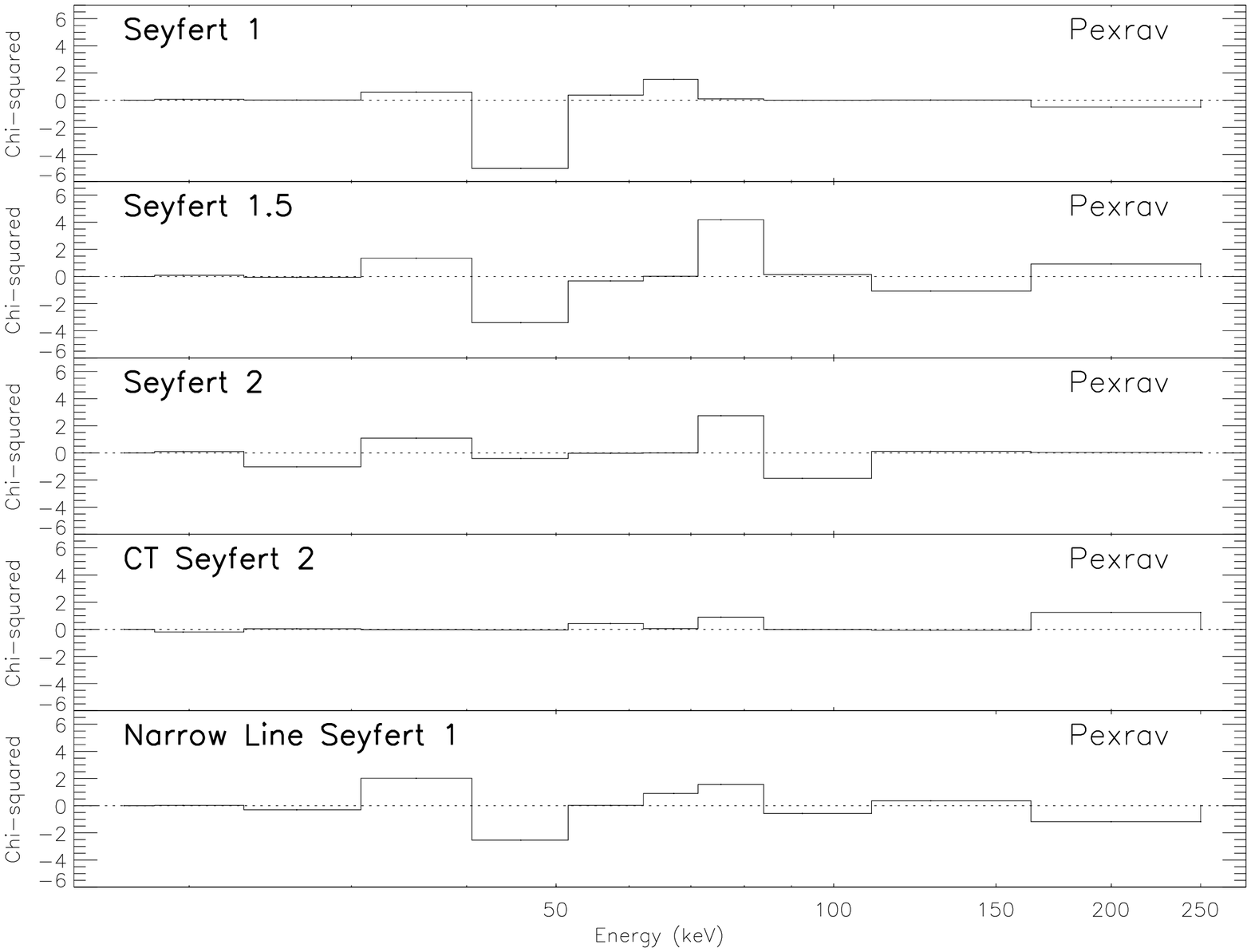}\end{minipage}
 \begin{minipage}[t]{1\textwidth}
  \caption{Spectra of different classes of AGN obtained using the $pexrav$ model. We also show the contributions to the $\chi^2$ relative to the best fits. The results of the spectral analysis are reported in Table\,\ref{tab:Sy_an}.}
\label{fig:spectra}
 \end{minipage}

\end{figure*}

\begin{figure*}[]
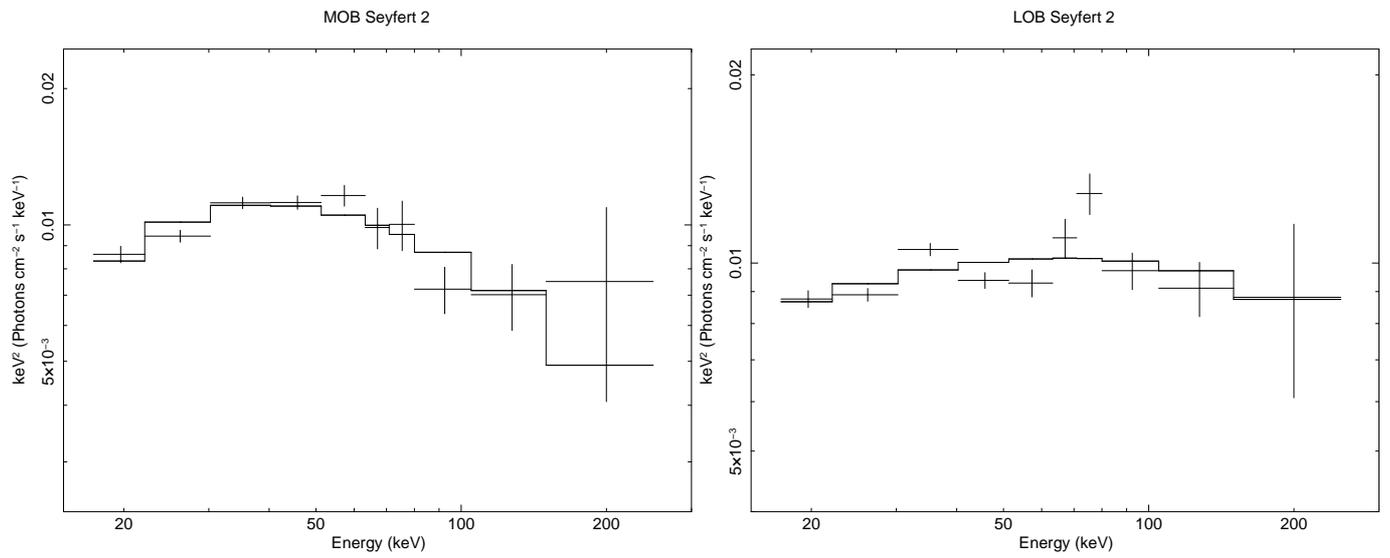

\centering
\begin{minipage}[!b]{.48\textwidth}
\centering
\includegraphics[height=9cm,angle=270]{16409_Msp.eps}
\end{minipage}
\hspace{0.05cm}
\begin{minipage}[!b]{.48\textwidth}
\centering
\includegraphics[height=9cm,angle=270]{16409_Lsp.eps} \end{minipage}
 \begin{minipage}[t]{1\textwidth}
  \caption{Spectra of MOB ({\it left panel}) and LOB ({\it right panel}) Seyfert\,2s, the model used is a cutoff power law plus a reflection component from neutral matter ($pexrav$). The results of the analysis are reported in Table\,\ref{tab:Sy_an_mob}.}
\label{fig:obsc_unobsc_spec}
 \end{minipage}
\end{figure*}

In Fig.\,\ref{fig:spectra}, we show the average 17--250\,keV spectra of Seyfert\,1s, Seyfert\,1.5s, Seyfert\,2s, CT Seyfert\,2s, and NLS1s. The parameters of the models used for the fit are reported in Table\,\ref{tab:Sy_an}.
In Fig\,\ref{fig:obsc_unobsc_spec}, we show the average spectra of MOB and LOB Seyfert\,2s. The parameters of the models are reported in Table\,\ref{tab:Sy_an_mob}.

\begin{acknowledgements}
We thank Marc T\"urler for providing us the {\it INTEGRAL} IBIS/ISGRI CXB spectrum, Carlo Ferrigno for his helpful comments on this work, and Piotr Lubinski for his help. We also thank the anonymous referee for his helpful comments. This research has made use of the NASA/IPAC Extragalactic Database (NED) which is operated by the Jet Propulsion Laboratory, of the SIMBAD Astronomical Database, which is operated by the Centre de Donn\'ees astronomiques de Strasbourg, and of the IGR Sources page maintained by J. Rodriguez \& A. Bodaghee \footnote{http://irfu.cea.fr/Sap/IGR-Sources/}.
\end{acknowledgements}

\textbf{References}
\renewcommand{\baselinestretch}{1.0}
\newenvironment{references}{
    \begin{list}{}{\leftmargin 1.5em
                                           \itemindent -1.5em
                                           \itemsep 0pt
                                           \parsep 0.1cm
                                           \footnotesize
                                            }
                                           }{\end{list} }
\newcommand{\entry}{\item}

\end{document}